\documentstyle[11pt,amssymb]{article}

\makeatletter
\@addtoreset{equation}{section}
\makeatother

\def\dalemb#1#2{{\vbox{\hrule height .#2pt
        \hbox{\vrule width.#2pt height#1pt \kern#1pt
                \vrule width.#2pt}
        \hrule height.#2pt}}}
\def\square{\mathord{\dalemb{6.8}{7}\hbox{\hskip1pt}}}

\def\0{{\sst{(0)}}}
\def\1{{\sst{(1)}}}
\def\2{{\sst{(2)}}}
\def\3{{\sst{(3)}}}
\def\4{{\sst{(4)}}}
\def\5{{\sst{(5)}}}
\def\6{{\sst{(6)}}}
\def\7{{\sst{(7)}}}
\def\8{{\sst{(8)}}}

\def\Z{\rlap{\sf Z}\mkern3mu{\sf Z}}
\def\R{\rlap{\rm I}\mkern3mu{\rm R}}

\def\ep{\epsilon}
\def\td{\tilde}
\def\wtd{\widetilde}

\def\crampest{\medmuskip = 1mu plus 1mu minus 1mu}
\def\uncramp{\medmuskip = 4mu plus 2mu minus 4mu}

\let\a=\alpha \let\b=\beta   \let\e=\epsilon

\let\C=\Chi

\def\nn{\nonumber} \def\bd{\begin{document}} \def\ed{\end{document}}
\def\ds{\documentstyle} \let\fr=\frac \let\bl=\bigl \let\br=\bigr
\let\Br=\Bigr \let\Bl=\Bigl
\let\bm=\bibitem
\let\na=\nabla
\let\pa=\partial \let\ov=\overline
\newcommand{\be}{\begin{equation}}
\newcommand{\ee}{\end{equation}}
\def\ba{\begin{array}}
\def\ea{\end{array}}
\def\ft#1#2{{\textstyle{{\scriptstyle #1}\over {\scriptstyle #2}}}}
\def\fft#1#2{{#1 \over #2}}
\def\del{\partial}
\def\sst#1{{\scriptscriptstyle #1}}
\def\oneone{\rlap 1\mkern4mu{\rm l}}
\def\ie{{\it i.e.\ }}
\def\via{{\it via}}
\def\semi{{\ltimes}}
\def\str{{\rm str}}
\def\jm{{\rm j}}
\def\im{{\rm i}}
\def\bOmega{{{\bar\Omega}}}
\def\Qn{{{Q_{\sst{\rm N}}}}}
\def\tX{{{\wtd X}}}
\def\C{{{\Bbb C}}}
\def\CP{{{\Bbb C}{\Bbb P}}}

\def\mapright#1{\smash{\mathop{-\!\!\!-\!\!\!-\!\!\!-\!\!\!-\!\!\!
             \longrightarrow}\limits^{#1}}}
\def\maprightt#1#2{\smash{\mathop{-\!\!\!-\!\!\!-\!\!\!-\!\!\!-\!\!\!
             \longrightarrow}\limits^{#1}_{#2}}}

\newcommand{\ho}[1]{$\, ^{#1}$}
\newcommand{\hoch}[1]{$\, ^{#1}$}
\newcommand{\bea}{\begin{eqnarray}}
\newcommand{\eea}{\end{eqnarray}}
\newcommand{\ra}{\rightarrow}
\newcommand{\lra}{\longrightarrow}
\newcommand{\Lra}{\Leftrightarrow}
\newcommand{\ap}{\alpha^\prime}
\newcommand{\bp}{\tilde \beta^\prime}
\newcommand{\tr}{{\rm tr} }
\newcommand{\Tr}{{\rm Tr} }

\newcommand{\NP}{Nucl. Phys. }
\newcommand{\tamphys}{\it Center for Theoretical Physics\\
Texas A\&M University, College Station, TX 77843}
\newcommand{\umich}{\it Department of Physics\\
University of Michigan, Ann Arbor, Michigan 48109}
\newcommand{\upenn}{\it Department of Physics and Astronomy\\
University of Pennsylvania, Philadelphia,  PA 19104}
\newcommand{\SISSA}{\it  SISSA-ISAS and INFN, Sezione di Trieste\\
Via Beirut 2-4, I-34013, Trieste, Italy}

\newcommand{\ihp}{\it Institut Henri Poincar\'e\\
  11 rue Pierre et Marie Curie, F 75231 Paris Cedex 05}

\newcommand{\damtp}{\it DAMTP, Centre for Mathematical Sciences,
 Cambridge University, Wilberforce Road, Cambridge CB3 OWA, UK}

\newcommand{\auth}{M. Cveti\v{c}\hoch{\dagger}, G.W. Gibbons\hoch{\sharp}, 
H. L\"u\hoch{\star} and C.N. Pope\hoch{\ddagger}}

\begin{document}
\begin{flushright}
\hfill{DAMTP-2000-131}\ \ \ {CTP TAMU-37/00}\ \ \ {UPR-911-T}\ \ \
{MCTP-00-08}\ \ \ {RUNHETC-2000-44}\ \ \ IHP-2000/07\ \ \
{hep-th/0012011}\ \ \
{December, 2000}
\end{flushright}


\begin{center}
{ \large {\bf Ricci-flat Metrics, Harmonic Forms and Brane Resolutions}}

\vspace{10pt}
\auth

\vspace{5pt}
{\hoch{\dagger}\upenn}

\vspace{5pt}
{\hoch{\sharp}\damtp}

\vspace{5pt}
{\hoch{\dagger} \it Department of Physics and Astronomy, Rutgers University,
Piscataway, NJ 08855}

\vspace{5pt}
{\hoch{\star}\umich}

\vspace{5pt}
{\hoch{\ddagger}\tamphys}

\vspace{5pt}
{\hoch{\ddagger,\dagger}\ihp}

\vspace{10pt}

\underline{ABSTRACT}
\end{center}

          We discuss the geometry and topology of the complete,
non-compact, Ricci-flat Stenzel metric, on the tangent bundle of
$S^{n+1}$.  We obtain explicit results for all the metrics, and show
how they can be obtained from first-order equations derivable from a
superpotential.  We then provide an explicit construction for the
harmonic self-dual $(p,q)$-forms in the middle dimension $p+q=(n+1)$
for the Stenzel metrics in $2(n+1)$ dimensions.  Only the
$(p,p)$-forms are $L^2$-normalisable, while for $(p,q)$-forms the
degree of divergence grows with $|p-q|$.  We also construct a set of
Ricci-flat metrics whose level surfaces are $U(1)$ bundles over a
product of $N$ Einstein-K\"ahler manifolds, and we construct examples
of harmonic forms there.  As an application, we construct new examples
of deformed supersymmetric non-singular M2-branes with such
8-dimensional transverse Ricci-flat spaces.  We show explicitly that
the fractional D3-branes on the 6-dimensional Stenzel metric found by
Klebanov and Strassler is supported by a pure $(2,1)$-form, and thus
it is supersymmetric, while the example of Pando Zayas-Tseytlin is
supported by a mixture of $(1,2)$ and $(2,1)$ forms.  We comment on
the implications for the corresponding dual field theories of our
resolved brane solutions.


\pagebreak
\setcounter{page}{1}

\tableofcontents
\vfill\eject

\section{Introduction}

    Fractional D3-branes have been extensively studied recently, since
they can provide supergravity solutions that are dual to
four-dimensional $N=1$ super-Yang-Mills theories in the infra-red regime
\cite{klebtsey,klebstra,ganpol,gub,zaytse,clpres,bgz,bvflmp}.  The idea is
that by turning on fluxes for the R-R and NS-NS 3-form fields of the
type IIB supergravity, in addition to the usual flux for the self-dual
5-form that supports the ordinary D3-brane, a deformed solution can be
found that is free of the usual small-distance singular behaviour on the
D3-brane horizon.  This is achieved by first replacing the usual flat
6-metric transverse to the D3-brane by a non-compact Ricci-flat K\"ahler
metric.  It can then be shown that if there exists a suitable harmonic
3-form $G_\3$ satisfying a complex self-duality condition, then the type
IIB equations of motion are satisfied if the R-R and NS-NS fields are
set equal to the real and imaginary parts of the harmonic 3-form, with
the usual harmonic function $H$ of the D3-brane solution now satisfying
the modified equation $\square \, H = -\ft1{12} m^2\, |G_\3|^2$ in the
transverse space.  A key feature of the type IIB equations that allows
such a solution to arise is that there is a Chern-Simons or
``transgression'' modification in the Bianchi identity for the self-dual
5-form, bilinear in the R-R and NS-NS 3-forms.

   The construction can be extended to encompass other examples of
$p$-brane solutions, and in \cite{clpres} a variety of such cases were
analysed.  These included heterotic 5-branes, dyonic strings, M2-branes,
D2-branes, D4-branes and type IIA and type IIB strings.  The case of
M2-branes was also discussed in \cite{hawtay}.  In all these cases, the
ability to construct deformed solutions depends again upon the existence
of certain Chern-Simons or transgression terms in Bianchi identities or
equations of motion.  The additional field strength contribution
that modifies the
standard $p$-brane configuration then comes from an appropriate harmonic
form in the transverse space.  One again replaces the usual flat
transverse space by a more general complete non-compact Ricci-flat
manifold.  In order to get deformed solutions that are still
supersymmetric, a necessary condition on this manifold is that it must
have an appropriate special holonomy that admits the existence of
covariantly-constant spinors.

    One can easily establish that if the harmonic form is
$L^2$-normalisable, then it is possible to choose integration constants
in such a way that the deformed solution is completely non-singular
\cite{clpres}.  In particular, it can be arranged that the horizon is
completely eliminated, with the metric instead smoothly approaching a
regular ``endpoint'' at small radial distances.  At large distances, the
metric then has the same type of asymptotic structure as in the
undeformed case, with a well-defined ADM mass per unit spatial
world-volume.  If, on the other hand, the harmonic form in the
transverse manifold is not $L^2$-normalisable, then the deformed
solution will suffer from some kind of pathology.  Usually, one chooses
a harmonic form that is at least square-integrable in the small-radius
regime, and this can be sufficient to allow a solution which gives a
useful infra-red description of the dual super-Yang-Mills theory.
 
    If the harmonic form fails to be square-integrable at large
radius, then this will lead to some degree of pathology in the
asymptotic structure of the deformed solution in that region.  For
example, the deformed KS D3-brane solution \cite{klebstra} is based on
a non-normalisable harmonic 3-form in the six-dimensional Ricci-flat
K\"ahler transverse space, for which the integral of $|G_\3|^2$
diverges as the logarithm of the proper distance at large radius.
This leads to a deformed D3-brane metric that is complete and
everywhere non-singular, and for which the harmonic
function $H$ has the asymptotic structure
\be
H\sim c_0 + \fft{Q + m^2\, \log \rho}{\rho^4}\label{d3asymp}
\ee
at large proper distance $\rho$.  Although the metric is still
asymptotic to $dx^\mu\, dx_\mu + ds_c^2$, where $ds_c^2$ is the metric
on the six-dimensional Ricci-flat conifold, the effect of the
deformation involving the logarithm is that the associated ADM mass
per unit 3-volume is no longer well-defined.  This is because the
effect of the $\log \rho$ term in $H$ is to cause a slower fall-off at
infinity than the normal $\rho^{-4}$ dependence that picks up a finite
and non-zero ADM contribution.\footnote{For practical purposes, the
ADM mass measured relative to the fiducial metric $dx^\mu\, dx_\mu +
ds_c^2$ is a certain constant times the limit of $\rho^5\, \del
H/\del\rho$ as $\rho$ goes to infinity.}  This change in the
asymptotic structure implies that the solution may not admit an
AdS$_5$ region, even when the constant $c_0$ in (\ref{d3asymp}) goes
to zero in a decoupling limit.  Of course this feature is itself of
great interest, since it is associated with a breaking of conformal
symmetry in the dual field theory picture.

   One might wonder whether there could be some other Ricci-flat
K\"ahler 6-manifold for which an $L^2$-normalisable harmonic 3-form
might exist.  In fact rather general arguments establish that this is
not possible, at least for the case where the 6-metric is
asymptotically of the form of a cone, and the middle homology is
one-dimensional.\footnote{We are grateful to
Nigel Hitchin for extensive discussions on this point.}  On the other
hand, $L^2$-normalisable harmonic forms can exist in non-compact
Ricci-flat manifolds in other dimensions, and indeed some examples of
fully resolved $p$-brane solutions based on such harmonic forms were
obtained in \cite{clpres}.  We shall obtain further examples in this
paper, using Ricci-flat K\"ahler 8-manifolds to obtain smooth
deformed M2-branes. Since the ADM mass is then well-defined, the
asymptotic structure correspondingly may still allow an approach to
AdS, if the constant term in the metric function $H$ goes to zero,
implying that the dual field theory will still be a conformal one
(three-dimensional in the case of M2-branes).

    In this paper, we explore some of these questions in greater detail.
To begin, in section 2, we study the class of complete non-compact
Ricci-flat K\"ahler manifolds whose metrics were constructed by Stenzel
\cite{sten}.  These are asymptotically conical, with level surfaces that
are described by the coset space $SO(n+2)/SO(n)$, and they have real
dimension $d=2n+2$.  The $n=1$ example is the Eguchi-Hanson instanton
\cite{egha}, and the $n=2$ example is the six-dimensional ``deformed
conifold'' found by Candelas and de la Ossa \cite{candel}.  It is this
example that is used in the fractional D3-brane KS solution in
\cite{klebstra}.  In section 2.1 we describe the geometry and topology
of the general Stenzel manifolds, and then in section 2.2 we carry out
detailed calculations of the curvature, and show how Ricci-flat
solutions can be obtained from a system of first-order equations
derivable from a superpotential.  In subsequent subsections we then
obtain the explicit Ricci-flat Stenzel metrics and their K\"ahler forms,
and then we derive integrability conditions for the covariantly-constant
spinors.

    In section 3 we obtain explicit results for harmonic forms in the
middle dimension, that is to say, for harmonic $(n+1)$-forms in the
$2(n+1)$-dimensional Stenzel metrics.\footnote{Nigel Hitchin has
informed us that Daryl Noyce has independently constructed the unique
harmonic form in the middle dimension in the $4N$-dimensional Stenzel
manifolds.}  More precisely, we construct harmonic $(p,q)$-forms for all
integers $p$ and $q$ satisfying $p+q=n+1$, where $p$ and $q$ count the
number of holomorphic and antiholomorphic indices.  We show that these
are $L^2$-normalisable if and only if $p=q$, which can, of course, occur
only in dimensions $d=4p$.

    In section 4, we make use of some of these results in order to
construct deformed $p$-brane solutions.  Specifically, we
first review the fractional D3-brane solution of \cite{klebstra}.  Our
results on harmonic forms allow us to give an explicit proof that their
solution has a harmonic 3-form of type $(2,1)$, which therefore ensures
supersymmetry.  We then construct a smooth deformed M2-brane, using
the $L^2$-normalisable $(2,2)$-form in the 8-dimensional Stenzel metric.
This is also supersymmetric.

   In section 5 we construct another class of complete non-compact
Ricci-flat K\"ahler manifolds.  These are again of the form of
resolved cones, but in this case the level surfaces are themselves
$U(1)$ bundles over the product of $N$ Einstein-K\"ahler manifolds.
Typical examples would be to take the base space to be ${\cal M}=
\prod_{i=1}^N \CP^{m_i}$, for an arbitrary set of integers $m_i$.  In
fact the requirements of regularity of the metric mean that one of the
factors in the base space ${\cal M}$ must be a complex projective
space, but the others might be other Einstein-K\"ahler manifolds.
Topologically, the total space is a $\C^k$ bundle over the remaining
Einstein-K\"ahler factors.\footnote{There are certain topological
restrictions on the possible choices for the other Einstein-K\"ahler
factors in the base space.  For a detailed discussion, see
\cite{d2frac}.}   Having obtained general results for
Ricci-flat K\"ahler metrics in all the cases, we present some more
detailed explicit formulae for three 8-dimensional examples,
corresponding to taking the base space to be $S^2\times \CP^2$,
$\CP^2\times S^2$ and $S^2\times S^2\times S^2$.  We also discuss some
well-known examples corresponding to complex line bundles over
$\CP^m$. 

   In section 6 we make use of our results for these Ricci-flat
metrics, to obtain further examples of deformed $p$-brane solutions.
We begin by considering the case where the base space is ${\cal
M}=S^2\times S^2$ (\ie $m_1=m_2=1$), meaning that the level surfaces
are the 5-dimensional space known as $T^{1,1}$ or $Q(1,1)$, which is a
$U(1)$ bundle over $S^2\times S^2$.  Topologically, the 6-dimensional
manifold is a $\C^2$ bundle over $\CP^1$.  Its Ricci-flat metric is
present in \cite{candel}, and it was discussed recently in
\cite{zaytse}, where it was used to provide an alternative resolution
of the D3-brane.  We construct the self-dual harmonic 3-form that was
used in \cite{zaytse} in a complex basis, and by this means
demonstrate that it contains both $(2,1)$ and $(1,2)$ pieces.  This
implies that the resolved D3-brane solution of \cite{zaytse} is not
supersymmetric \cite{clpres}.  We also construct $L^2$-normalisable
harmonic 4-forms of type $(2,2)$ in the 8-dimensional examples based
on $S^2\times \CP^2$ and $S^2\times S^2\times S^2$, and then use these
in order to construct additional deformed M2-branes, which
are supersymmetric.  A further smooth deformed M2-brane example,
which is non-supersymmetric, results from taking the 8-dimensional
transverse space to be the complex line bundle over $\CP^3$.  We also
include a discussion of a fifth completely smooth deformed M2-brane,
which was obtained previously in \cite{clpres}.  This solution uses an
8-manifold of exceptional Spin(7) holonomy rather than a Ricci-flat
K\"ahler manifold.  We give a simple proof of its supersymmetry.

   The paper ends with conclusions and discussions in section 7.

\section{Stenzel metrics}

    In this section we shall construct a sequence of complete
non-singular Ricci-flat K\"ahler metrics, one for each even dimension,
on the co-tangent bundle of the $(n+1)$ sphere $T^\star S^{n+1}$.
Restricted to the base space $S^{n+1}$, the metric coincides with the
standard round sphere metric.  The sequence, which begins with the
Eguchi-Hanson metric for $n=1$, was first constructed in generality by
Stenzel \cite{sten} following a method discussed in \cite{GP}. The case
$n=2$ was originally given, in rather different guise, by Candelas and
de la Ossa \cite{candel} as a ``deformation" of the conifold.  The
isometry group of these metrics is $SO(n+2)$, acting in the obvious way
on $T^\star S^{n+1}$.  The principal (\ie generic) orbits are of
co-dimension one, corresponding to the coset $SO(n+2)/SO(n)$.  There is
a degenerate orbit (\ie a generalized ``bolt'') corresponding to the
zero section, \ie to the base space $S^{n+1}\equiv SO(n+2)/SO(n+1)$.  It
is therefore possible to obtain the ordinary differential equations
satisfied by the metric functions using coset techniques, and this we
shall do shortly.  Before doing so, however, we wish to make some
comments about the geometry and topology of the metrics, which are
intended to illuminate the subsequent calculations.

\subsection{Geometrical and topological considerations} 

   Any K\"ahler metric is necessarily symplectic, and in the present
case the symplectic structure coincides with the standard symplectic
structure on $T^\star S^{n+1}$. The sphere $S^{n+1}$ is thus
automatically a Lagrangian sub-manifold. In other words the K\"ahler
form restricted to the $(n+1)$-sphere vanishes. The complex structure on
$T^\star S^{n+1}$ is however non-obvious, and arises from the fact that
we may view $T^\star S^{n+1}$ as a complex quadric in ${\Bbb C}^{n+2}$,
\be
z^a \, z^a=a^2,
\ee
where $ a= 1,2,\dots ,n+2$. Setting
\be
z^a= \cosh (\sqrt{p^b\, p^b} )\,  x^a+ \im\,
{\sinh (\sqrt{p^b\, p^b}) \over \sqrt{p_b \, p_b}} \, p_a,
\ee
one obtains $x^b \, x^b= a^2 $ and $p_b\, x^b=0$.  These are the
equations defining a point $x^b$ lying on an $(n+1)$-sphere of radius
$a$ in ${\Bbb E}^{n+1}$, and a cotangent vector $p_b$. Note that as the
radius $a$ is sent to zero we obtain the conifold, which makes contact
with the work of Candelas and de la Ossa \cite{candel}.

    The strategy of Stenzel \cite{sten} is now to assume that the
K\"ahler potential $K$ depends only on the quantity
\be
\tau= {\bar z}^a\, z^a =\cosh(2\sqrt{p_b \, p_b}).
\ee
From this it is clear that the principal orbits of the isometry group
correspond to the surfaces of constant energy $H=\ft12 p_b \, p_b $ on
the phase space $ T^\star S^{n+1}$. The stabliser of each point on the
orbit consists of rotations leaving fixed a point on $S^{n+1}$ and a
tangent vector $p_b$. The transitivity of the action is equally obvious.
Thus $\sqrt{p_b\, p_b}$, or some function of it, it will serve as a
radial variable.

   In fact the level sets $H={\rm constant}$ can be viewed as circle
bundles over the Grassmannian $SO(n+2)/(SO(n) \times SO(2))$.  To see
why, recall that the Hamiltonian $H$ generates the geodesic flow on
$T^\star S^{n+1}$.  Each such geodesic is a great circle consisting of
the intersection of a two-plane through the origin of ${\Bbb E}^{n+2}$
with the $(n+1)$-sphere.  The circle factor in the denominator of the
coset corresponds to the fact that geodesics or great circles are the
orbits of a circle subgroup of the isometry group $SO(n+2)$ of the
$(n+1)$-sphere.

   Thus the circle fibre of the circle bundle is an orbit of the
isometry group of the Ricci-flat K\"ahler metric.  In terms of K\"ahler
geometry, the quotient of $T^\star S^{n+1}$ by the circle action
corresponds to the Marsden-Weinstein or symplectic quotient, and gives
at each radius a homogeneous K\"ahler metric of two less dimensions.

    At large distances the Stenzel metric tends to a Ricci-flat cone
over the Einstein-Sasaski manifold $SO(n+2) /SO(n)$.  At small radius
the orbits collapse to the zero-section of $T^\star S^{n+1}$.  Thus it
is clear that the $(n+1)$-sphere $\Sigma \in H_{n+1}( T^\star S^{n+1})$
provides the only interesting homology cycle, and it is in the middle
dimension.  In the case that $n$ is odd, its self-intersection number
$\Sigma \cdot \Sigma \in {\Bbb Z}$ is, depending upon orientation
convention, 2, while if $n$ is even its self-intersection number
vanishes.  This is equivalent to the statement that the Euler
characteristic of the even-dimensional spheres is 2, while for the
odd-dimensional spheres it vanishes.  To see this equivalence, recall
that the topology of the co-tangent bundle is the same as that of the
tangent bundle. Now the Euler characteristic of any closed orientable
manifold is given by the number of intersections, suitably counted, of
the zero section with any other section of its tangent bundle.  In other
words it is the number of zeros, suitably counted, of a vector field on
the manifold.

   We shall see that these facts have consequences for the cohomology.
In the case of a closed $(2n+2)$-manifold ${\cal M}$ (\ie compact,
without boundary), one may use Poincar\'e duality to see that if
$\alpha$ and $\beta$ are closed middle-dimensional $(n+1)$-forms
representing elements of $H^{n+1}({\cal M})$, then the cup product
$\alpha \cup \beta$ is an integer-valued bilinear form on $H^{n+1}
({\cal M})$ given by
\be
\int_{\cal M} \alpha \wedge \beta\,.\label{cup1}
\ee
The cup product is symmetric or skew-symmetric depending upon whether
$n$ is odd or even respectively.   Thus if $n$  is even,
\be
\int _{\cal M} \alpha \wedge \alpha =0\,.
\ee
Moreover, the Hodge duality operator $\star $ acts on $H^{n+1} ({\cal
M})$, and
\be
\star \star = (-1) ^{n+1}\,.
\ee
Thus if $n$ is odd, $H^{n+1}({\cal M})$ decomposes into real self-dual
or anti-self dual $(n+1)$ forms. Any such closed form must necessarily
be harmonic, and its $L^2$ norm will be proportional to the
self-intersection number. The total number of linearly-independent
harmonic middle-dimensional forms will depend only on the topology of
the closed manifold ${\cal M}$.

    If $n$ is even, we can find a complex basis of self-dual harmonic
forms in $L^2$, but there is no relation between their normalisability
and the integral in (\ref{cup1}).

   Our manifolds are non-compact, and the situation is therefore more
complicated and we must proceed with caution.  The usual one-to-one
correspondence between harmonic forms and geometric cycles may break
down.  One generally expects at least as many $L^2$ harmonic forms as
topology requires, but there may be more (c.f. \cite{SegalSelby}).  It
is still true that $L^2$ harmonic forms must be closed and co-closed
\cite{Rham}. However, the notion of exactness must be modified since we
are interested in whether closed forms in $L^2$ are the exterior
derivatives of forms of one lower degree which are also in $L^2$.  For
example, the Taub-NUT metric admits an exact harmonic two-form in $L^2$,
but it is the exterior derivative of a Killing 1-form which is not in
$L^2$.

   In the present case, if $n$ is odd it seems reasonable to expect at
least one harmonic form in the middle dimension, which is Poincar\'e
dual to the $(n+1)$-sphere. Because the Stenzel metric behaves like a
cone near infinity, all the Killing vectors are of linear growth. It
follows \cite{Hitchin} that any harmonic form must be invariant under
the action of the isometry group.  In the case of the Taub-NUT and
Schwarzschild metrics, this observation permits the complete
determination of the $L^2$ cohomology \cite{Hitchin,etehau}.  We shall
obtain an $L^2$ harmonic form in the middle dimension for all the
Stenzel manifolds with odd $n$.

    We obtain a general explicit construction of harmonic $(p,q)$-forms
in all the Stenzel manifolds, where $p+q=n+1$.  These middle-dimension
harmonic forms include $(p,p)$ forms when $n$ is odd, and these are the
$L^2$-normalisable examples mentioned above.  All the others are
non-normalisable, with a ``degree of non-normalisability'' that
increases with $|p-q|$ at fixed $p+q$.  In particular, this accords with
the expectation that if $n$ is even we should not find any harmonic form
in $L^2$.\footnote{ Nigel Hitchin and  Tamas Hausel have both pointed out to
us that results of Atiyah, Patodi and Singer on asyptotically cylindrical
manifolds \cite{atipatsin} and some
propeties of K\"ahler manifolds used in \cite{Hitchin} can be
extended to asymptotically conical metrics,
and they imply that the $L^2$ cohomology is toplogical, \ie isomorphic
to the compactly-supported cohomolgy in ordinary cohomolgy.
The results reported here are consistent with those theorems.
We thank them for helpful communications.}

\subsection{Detailed calculations}

    Let $L_{AB}$ be the left-invariant 1-forms on the group manifold
$SO(n+2)$.  These satisfy
\be
dL_{AB} = L_{AC} \wedge L_{CB}\,.
\ee
We consider the $SO(n)$ subgroup, by splitting the index as
$A=(1,2,i)$.  The $L_{ij}$ are the left-invariant 1-forms for the
$SO(n)$ subgroup.  We make the following definitions:
\be
\sigma_i \equiv L_{1i}\,,\qquad \td\sigma_i \equiv L_{2i}\,,\qquad
\nu \equiv L_{12}\,.
\ee
These are the 1-forms in the coset $SO(n+2)/SO(n)$.  We have
\bea
&&d\sigma_i = \nu\wedge \td\sigma_i + L_{ij}\wedge \sigma_j\,,\quad
d\td\sigma_i = -\nu\wedge \sigma_i + L_{ij}\wedge
\td\sigma_j\,,\quad
d\nu = -\sigma_i\wedge \td\sigma_i\,,\nn\\
&&dL_{ij} = L_{ik}\wedge L_{kj} -\sigma_i\wedge \sigma_j -
\td\sigma_i\wedge \td\sigma_j\,.\label{exd}
\eea
Note that the 1-forms $L_{ij}$ lie outside the coset, and so one finds
that they do not appear eventually in the expressions for the
curvature (see also \cite{danstr}).

   We now consider the metric
\be
ds^2 = dt^2 + a^2 \sigma_i^2 + b^2\, \td\sigma_i^2 + c^2\, \nu^2\,,
\ee
where $a$, $b$ and $c$ are functions of the radial coordinate $t$, and
then we define the vielbeins
\be
e^0=dt\,,\qquad e^i = a\, \sigma_i\,,\qquad e^{\td i} = b\,
\td\sigma_i\,,\qquad e^{\td 0}=c\, \nu\,.\label{vielbein}
\ee
Calculating the spin connection, we find
\bea
&&\omega_{0i} = -\fft{\dot a}{a}\, e^i\,,\qquad
\omega_{0\td i} = -\fft{\dot b}{b}\, e^{\td i}\,,\qquad
\omega_{0\td 0} = -\fft{\dot c}{c}\, e^{\td 0}\,,\nn\\
&&\omega_{\td 0i} =  B\, e^{\td i} \,,\qquad
\omega_{\td 0\td i} = - A\, e^{i} \,,\qquad          
\omega_{i\td j} =  C\, \delta_{ij}\, e^{\td 0}\,,\nn\\
&&\omega_{ij} = -L_{ij}\,,\qquad \omega_{\td i \td j} = - L_{ij}\,,
\label{spincon}
\eea
where a dot means $d/dt$, and
\be
 A = \fft{(a^2-b^2-c^2)}{2 a\, b\, c}\,,\qquad
 B = \fft{(b^2-c^2-a^2)}{2 a\, b\, c}\,,\qquad
 C = \fft{(c^2-a^2-b^2)}{2 a\, b\, c}\,.\label{abcdef}
\ee

   From this, we obtain the curvature 2-forms
\bea
\Theta_{0i} &=&  -\fft{\ddot a}{a}\, e^0\wedge e^i - \Big( \fft{\dot
a}{b\, c} + \fft{ C\, \dot b}{b} + \fft{ B\, \dot c}{c}\Big)\,
e^{\td 0} \wedge e^{\td i}\,,\nn\\
\Theta_{0\td i} &=& -\fft{\ddot b}{b}\, e^0\wedge e^{\td i} +
\Big( \fft{\dot
b}{a\, c} + \fft{ C\, \dot a}{a} + \fft{ A\, \dot c}{c}\Big)\,
e^{\td 0} \wedge e^{i}\,,\nn\\
\Theta_{0\td 0} &=& -\fft{\ddot c}{c}\, e^0\wedge e^{\td 0} +
\Big( \fft{\dot
c}{a\, b} + \fft{ B\, \dot a}{a} + \fft{ A\, \dot b}{b}\Big)\,
e^i\wedge e^{\td i}\,,\nn\\
\Theta_{ij} &=& \Big(\fft1{a^2} - \fft{\dot a^2}{a^2}\Big)\, e^i\wedge
e^j + \Big( \fft1{b^2} - B^2\Big)\, e^{\td i}\wedge e^{\td
j}\,,\nn\\
\Theta_{\td i \td j} &=& \Big(\fft1{b^2} - \fft{\dot b^2}{b^2}\Big)\,
e^{\td i}\wedge e^{\td j} +
\Big( \fft1{a^2} - A^2\Big)\, e^{i}\wedge e^{ j}\,,\nn\\
\Theta_{i \td j} &=&  A\,  B \, e^{\td i}\wedge e^j -\fft{\dot a\,
\dot b}{a\, b}\, e^i\wedge e^{\td j} - \fft{ C\, c}{a\, b}\,
\delta_{ij}\, e^k\wedge e^{\td k} + \Big( \dot C + \fft{ C\,
\dot c}{c}\Big)\, \delta_{ij}\, e^0\wedge e^{\td 0}\,,\nn\\
\Theta_{\td 0i} &=& -\Big(\fft{\dot a\, \dot c}{a\, c} +  A\,  C +
        \fft{ B\, b}{a\, c}\Big)\, e^{\td 0}\wedge e^i +
   \Big( \dot B + \fft{ B\, \dot b}{b}\Big)\, e^0\wedge e^{\td
        i}\,,\nn\\
\Theta_{\td0\td i} &=& -\Big(\fft{\dot b\, \dot c}{b\, c} +  B\,  C +
        \fft{ A\, a}{b\, c}\Big)\, e^{\td 0}\wedge e^{\td i} -
   \Big( \dot A + \fft{ A\, \dot a}{a}\Big)\, e^0\wedge e^{i}\,.
\label{stenzelcurv}
\eea

   This implies that the Ricci tensor is diagonal, and that its
vielbein components are given by         
\bea
R_{00} &=& -\fft{n\, \ddot a}{a} - \fft{n\, \ddot b}{b} - \fft{\ddot
c}{c}\,,\nn\\
R_{\td0\td0} &= &- \fft{\ddot c}{c}   -n\, \Big( \fft{\dot a\, \dot c}{a\,
c}  + \fft{\dot b\, \dot
c}{b\, c} + \fft{(a^2-b^2)^2 -c^4}{2 a^2\, b^2\, c^2}\Big) \,,\nn\\
R_{ij} &=& \Big[ -\fft{\ddot a}{a} +(n-1)\,
     \Big(\fft1{a^2} - \fft{\dot a^2}{a^2} \Big) -
  \fft{n\, \dot a\, \dot b}{a\, b} - \fft{\dot a\, \dot c}{a\, c}
  + \fft{a^4-(b^2-c^2)^2}{2 a^2\, b^2\, c^2} \Big]\, \delta_{ij}\,,\nn\\
R_{\td i\td j} &=& \Big[ -\fft{\ddot b}{b} +(n-1)\,
     \Big(\fft1{b^2} - \fft{\dot b^2}{b^2} \Big) -
  \fft{n\, \dot a\, \dot b}{a\, b} - \fft{\dot b\, \dot c}{b\, c}
  + \fft{b^4-(a^2-c^2)^2}{2 a^2\, b^2\, c^2}\Big]\, 
\delta_{ij}\,,\label{stenzelricci}
\eea

    Defining $a=e^\a$, $b=e^\beta$, $c=e^\gamma$, and introducing the
new coordinate $\eta$ by $a^n\, b^n\, c\, d\eta = dt$, we find that the
Ricci-flat equations can be derived from the Lagrangian $L=T-V$, where
\bea
T &=& \a' \, \gamma'  
+ \beta'\,  \gamma'  
+ n\, \a'\, \beta' + \ft12(n-1)\, {\a'}^2 +\ft12 (n-1)\,{\beta'}^2 
\,,\nn\\
V&=& \ft14 (a\, b)^{2n-2}\, (a^4+b^4+c^4 - 2a^2\, b^2 -2n\, a^2\, c^2
-2n\, b^2\, c^2)\,,
\eea
where a prime means $d/d\eta$,  
together with the constraint that the Hamiltonian vanishes, $T+V=0$.
(Note that the Hamiltonian comes from the $G_{00}$ component
of the Einstein tensor.)

   Writing the Lagrangian as $L=\ft12 g_{ij}\, (d{\a^i}/d\eta)\, 
(d{\a^j}/d\eta) -V$,
where $\a^i=(\a,\beta,\gamma)$, 
we find that the potential can be written
in terms of a superpotential, as
\be
V = - \ft12 g^{ij} \, \fft{\del W}{\del \a^i}\, \fft{\del W}{\del \a^j}
\ee
with 
\be
W = \ft12 (a\, b)^{n-1}\, (a^2 + b^2 + c^2)\,.
\ee
It follows that the Lagrangian can be written, after dropping a total
derivative, as
\be
L = \ft12 g_{ij}\, \Big(\fft{d{\a^i}}{d\eta} \pm g^{ik}\, \del_k
W\Big) \, \Big(\fft{d{\a^j}}{d\eta} \pm 
             g^{j\ell}\, \del_\ell W\Big)\,,
\ee
where $\del_i W\equiv \del W/\del\a^i$.  This implies that the
second-order equations for Ricci-flatness are satisfied if the
first-order equations $d{\a^i}/d\eta = \mp g^{ij}\, \del_j W$ are satisfied.
Thus we arrive at the first-order equations
\bea
\dot \a &=& \ft12 e^{-\a-\beta-\gamma}\, (e^{2\beta} + e^{2\gamma}
-e^{2\a})\,,\nn\\
\dot \beta &=& \ft12 e^{-\a-\beta-\gamma}\, (e^{2\a} + e^{2\gamma}
-e^{2\beta})\,,\nn\\
\dot \gamma &=& \ft12 n\, e^{-\a-\beta-\gamma}\, (e^{2\a} + e^{2\beta}
-e^{2\gamma})\,,\label{firstorder}
\eea
where the dot again denotes the radial derivative $d/dt$.  Note that
in terms of the quantities defined in (\ref{abcdef}), these equations
take the simple form
\be
\dot\a +A=0\,,\qquad \dot\beta + B=0\,,\qquad \dot\gamma + n\, C=0\,.
\label{firstorder3}
\ee

    If we now make use of the first-order Ricci-flat conditions
(\ref{firstorder}) in the expressions (\ref{stenzelcurv}) for the
curvature 2-forms, we find that they can be simplified to
\crampest
\bea
&&\Theta_{0i} = -\fft{\ddot a}{a}\, (e^0\wedge e^i - e^{\td 0}\wedge e^{\td
i})\,,\qquad
\Theta_{0\td i} = -\fft{\ddot b}{b}\, (e^0\wedge e^{\td i} + e^{\td
0}\wedge e^i)\,,\nn\\
&&\Theta_{0\td 0} = -\fft{\ddot c}{c}\, (e^0 \wedge e^{\td 0} +
\fft1{n}\, e^i\wedge e^{\td i})\,,\qquad
\Theta_{ij} = \Big(\fft1{a^2} -\fft{\dot a^2}{a^2}\Big)\, 
(e^i\wedge e^j + e^{\td i}\wedge e^{\td j})\,,\nn\\
&&\Theta_{\td i \td j} = \Big(\fft1{b^2} -\fft{\dot b^2}{b^2}\Big)\, 
(e^i\wedge e^j + e^{\td i}\wedge e^{\td j})\,,\nn\\
&&\Theta_{i \td j} = A\, B\, (e^{\td i}\wedge e^j - e^i\wedge e^{\td
j} + \fft{2}{n}\, \delta_{ij}\, e^k\wedge e^{\td k}) - \Big(2 A\, B +
\fft{n\, C\, c}{a \, b}\Big)\, (e^0\wedge e^{\td 0} + \fft{1}{n}\,
e^k\wedge e^{\td k})\, \delta_{ij}\,,\nn\\
&&\Theta_{\td 0 i} = -\Big( \fft{\dot a\, \dot c}{a\, c} + A\, C +
\fft{B\, b}{a\, c}\Big)\, (e^{\td 0}\wedge e^i + e^0\wedge e^{\td i})
\,,\nn\\
&&\Theta_{\td 0 \td i} = -\Big( \fft{\dot b\, \dot c}{b\, c} + B\, C +
\fft{A\, a}{b\, c}\Big)\, (e^{\td 0}\wedge e^{\td i} - e^0\wedge e^{i})
\,,\label{stencurv2}
\eea
\uncramp

\subsection{Covariantly-constant spinors}

   Since the Stenzel metrics are K\"ahler, it follows that if they are 
Ricci flat then there should be two covariantly-constant spinors $\eta$.
The integrability condition is
\be
R_{abcd}\, \Gamma^{cd}\, \eta=0\,.
\ee
From the expressions for the curvature that we obtained in
(\ref{stencurv2}), 
we can then read off that the covariantly-constant spinors must satisfy 
\be
(\Gamma_{0i} - \Gamma_{\td 0 \td i})\, \eta =0\,,
\label{gammacon}
\ee
and it is easy to check that all the integrability conditions are
satisfied if (\ref{gammacon}) is satisfied. It is useful to note
that one can directly read off from (\ref{stencurv2}) other consequent
results (which can also be derived from (\ref{gammacon})), such as
$\Gamma_{ij}\, \eta = -\Gamma_{\td i\td j}\, \eta$.

    The covariant-constancy condition $D\, \eta \equiv d\, \eta +
\ft14 \omega_{ab}\, \Gamma^{ab}\, \eta=0$ itself is now easily solved.
From (\ref{spincon}), and using the first-order equations
(\ref{firstorder3}) and the integrability relations (\ref{gammacon}),
we find that $\eta$ simply satisfies $d\, \eta=0$.  In other words, in
the frame we are using the covariantly constant spinors have constant
components, and satisfy the projection conditions (\ref{gammacon}).
In fact we can reverse the logic, and {\it derive} the first-order equations  
(\ref{firstorder3}) by requiring the existence of a
covariantly-constant spinor $\eta$ subject to the additional
assumption that $\eta$ has constant components.   Equation
(\ref{gammacon}) is then a further consequence.  Moreover, the 2-form
\be
\bar \eta \, \Gamma_{ab}\, \eta\label{ebe}
\ee
is covariantly constant and may be normalised so that it squares to
$-1$; in other words, it gives us the K\"ahler form.

\subsection{K\"ahler form}

   From now on, we define a new radial coordinate $r$ related to $t$
by $dt= h\, dr$, where $h$ can be chosen for convenience, and a prime
will mean a derivative with respect to $r$.  Thus the metric is now
written as
\be
ds^2 = h^2\, dr^2 + a^2\, \sigma_i^2 + b^2\, \td\sigma_i^2 + c^2\,
\nu^2\,,\label{stenmeth}
\ee
and the vielbein is
\be
e^0=h\, dr\,,\qquad e^i = a\, \sigma_i\,,\qquad e^{\td i} = b\,
\td\sigma_i\,,\qquad e^{\td 0} = c\, \nu\,.\label{stenviel}
\ee

   It is easy to see that the K\"ahler form mentioned above is given by
\be
J = -e^0\wedge e^{\td 0} + e^i\wedge e^{\td i} = 
  -h\, c\, dr\wedge \nu + a\, b\, \sigma_i\wedge \td\sigma_i\,.
\label{stenkah}
\ee
The closure of $J$ follows from $(a \, b)' = h\, c$, which can be seen
from the first-order equations (\ref{firstorder}).   Further checking,
using the spin connection (\ref{spincon}), shows that $J$ is indeed
covariantly constant.  Again, the logic could be reversed, and by
requiring the existence of a covariantly-constant 2-form that squares
to $-1$, one could derive the first-order equations
(\ref{firstorder3}).  

   From this, it follows that we can introduce a holomorphic
tangent-space basis of complex 1-forms $\ep^\a$ as follows:
\be
\ep^0 \equiv -e^0 + \im\, e^{\td 0}\,,\qquad \ep^i = e^i + \im\,
e^{\td i}\,.\label{complexbasis}
\ee
In terms of this, we have that the K\"ahler form is
\be
J= \ft{\im}{2}\, \ep^\a\wedge \bar\ep^{\bar\a}\,,
\ee
and so it is manifestly of type $(1,1)$.  (One barred, one unbarred,
complex index.)

    By looking at how other forms are expressed in terms of the complex
holomorphic basis $\ep^\a$, we can see how they decompose into type
$(p,q)$ pieces, where $p$ and $q$ count the number of holomorphic and
anti-holomorphic basis 1-forms in each term.

\subsection{Explicit solutions for Ricci-flat Stenzel metrics}

    Here, we shall construct the explicit solutions to the first-order
equations (\ref{firstorder}), for arbitrary $n$.  This gives the class
of Ricci-flat metrics on complete non-compact manifolds of dimension
$d=2n+2$, as constructed by Stenzel.  Starting from (\ref{firstorder}),
and changing to the new radial coordinate $r$ related to $t$ by $dt=h\,
dr$, we first make the coordinate gauge choice $h=c$.  The first-order
equations then give
\bea
&&\a'-\b' = -2\sinh(\a-\beta)\,,\qquad \a'+\beta' =
e^{-\a-\beta+2\gamma}\,,\nn\\
&& \gamma' + \ft12 n\, (\a'+\beta' ) =n\,
\cosh(\a-\beta)\,.
\eea
The first equation gives $e^{\a-\beta}= \coth r$, the
third gives $e^{\a+\beta}= k\, e^{-2\gamma/n}\, \sinh 2r$, where $k$
is a constant, and then the second can be solved explicitly for
$\gamma$.  It is advantageous to introduce a function $R(r)$,
defined by
\be
R(r) \equiv  \int_0^r (\sinh 2u)^n\, du\,.\label{rdef}
\ee
Choosing $k=(n+1)^{-1/n}$ without loss of generality, the solution is
then given by
\bea
&&a^2\equiv e^{2\a} = R^{1/(n+1)}\, \coth r\,,\nn\\
&&b^2\equiv e^{2\beta} = R^{1/(n+1)}\, \tanh r\,,\\ 
&&
h^2=c^2\equiv e^{2\gamma} = \fft1{n+1}\, R^{-n/(n+1)}\, (\sinh
2r)^n\,,\nn
\eea
with the Ricci-flat metric taking the form
\be
ds^2 = c^2\, dr^2 + c^2\, \nu^2 + a^2\, \sigma_i^2 + b^2\,
\td\sigma_i^2\,.
\ee

    The integral (\ref{rdef}) can be evaluated in general, in terms of
a hypergeometric function:
\be
R= \fft{2^n}{n+1}\, (\sinh r)^{n+1}\,\, _2F_1\left[ \ft12(1+n),
\ft12(1-n), \ft12(3+n); -\sinh^2 r\right]\,.\label{rsol}
\ee
For each $n$ the result is expressible in relatively simple terms; for
the first few values of $n$ one has
\bea
n=1:&& R = \sinh^2 r\,,\nn\\
n=2:&& R =  \ft18(\sinh 4 r- 4r)\,,\nn\\
n=3:&& R = \ft23 (2+\cosh 2r)\, \sinh^4 r\,,\nn\\
n=4:&& R = \ft1{64} (24r - 8\sinh 4 r + \sinh 8r)\,,\nn\\
n=5:&& R = \ft2{15} (19 + 18 \cosh 2r + 3\cosh 4 r)\, \sinh^6 r\,,\nn\\
n=6:&& R = \ft1{384} (-120 r + 45 \sinh 4 r -9 \sinh 8r + \sinh 12 r)\,.
\eea
Note that when $n$ is odd, one can always change to a new radial
variable $z=\sinh r$ in terms of which the metric can be written using
rational functions.

   It is evident from (\ref{rsol}) that at small $r$ we shall have
\be
R \sim \fft{2^n}{n+1}\, r^{n+1}\,,
\ee
and consequently, the metric near $r=0$ takes the form
\be
ds^2 \sim \Big(\fft{2^n}{n+1}\Big)^{1/(n+1)}\, \Big[ dr^2 + r^2 \,
\td\sigma_i^2 + \sigma_i^2 + \nu^2\Big]\,.
\ee
Thus the radial coordinate runs from $r=0$, where the metric
approaches $\R^{n+1}\times S^{n+1}$ with an $S^{n+1}$ ``bolt,''
to the asymptotic region at $r=\infty$.  Note that the $S^{n+1}$ bolt
at $r=0$ is a Lagrangian submanifold; in other words, the K\"ahler
form (\ref{stenkah}) vanishes when restricted to it.

    When $n=1$, the 4-dimensional metric is the
Eguchi-Hanson instanton \cite{egha}.  When $n=2$, the 6-dimensional
metric is the ``deformed'' conifold solution found by Candelas and de la
Ossa \cite{candel}.  For arbitrary $n$, the solutions were first
obtained by Stenzel \cite{sten}.

\section{Harmonic forms}

\subsection{Harmonic $(p,q)$-forms in $2(p+q)$ dimensions}

    Here, we present a general construction of harmonic forms in the
``middle dimension,'' namely $(n+1)$-forms in the $2(n+1)$-dimensional
Stenzel manifolds.  These can be further refined as $(p,q)$ forms where
$p$ and $q$ denote the numbers of holomorphic and antiholomorphic
indices on the form, and $p+q=n+1$.

   We begin by making the following ansatz for the $(p,q)$ harmonic
form:
\bea
G_{\sst{(p,q)}} &=& f_1\,\ep_{i_1\cdots i_{q-1}j_1\cdots j_{p}}\,
\bar \ep^0\wedge \bar \ep^{i_1}\wedge\cdots \wedge\bar\ep^{i_{q-1}}\wedge
\ep^{j_1}\wedge\cdots\wedge\ep^{j_{p}}\nn\\
&&+f_2\,\ep_{i_1\cdots i_{p-1}j_1\cdots j_{q}}\,
\ep^0\wedge \ep^{i_1}\wedge\cdots\wedge \ep^{i_{p-1}}\wedge
\bar \ep^{j_1}\wedge\cdots\wedge \bar \ep^{j_q}\,,\label{pqans}
\eea   
where $f_1$ and $f_2$ are functions of $r$.  It is easy to see that the
epsilon tensors cause each term in each sum to be a product of complex
vielbeins in distinct subspaces each of complex dimension
one,\footnote{There are no factors such as $\ep^1\wedge \bar\ep^1$, for
example.  This also shows that these $(p,q)$-forms are entirely
perpendicular to the K\"ahler form $J=\ft{\im}{2} \, \ep^\a\wedge
\bar\ep^{\bar\a}$.}  and from this it follows that the Hodge dual is
given by
\be
{*G_{\sst{(p,q)}}} = \im^{\, p-q}\, G_{\sst{(p,q)}}\,.
\ee
Since $G_{\sst{(p,q)}}$ is an eigenstate under $*$, it follows that
the condition for harmonicity reduces to $dG_{\sst{(p,q)}}=0$.  

    It is useful first to note that from the expressions for the
vielbeins in the Stenzel metrics, we can rewrite (\ref{pqans}), up to
an irrelevant constant factor, as 
\bea
G_{\sst{(p,q)}} &=& f_1\,\ep_{i_1\cdots i_{q-1}j_1\cdots j_{p}}\,
(dr+\im\, \nu)\wedge \bar h^{i_1}\wedge\cdots \wedge\bar h^{i_{q-1}}\wedge
h^{j_1}\wedge\cdots\wedge h^{j_{p}}\nn\\
&&+f_2\,\ep_{i_1\cdots i_{p-1}j_1\cdots j_{q}}\,
(dr-\im \nu)\wedge h^{i_1}\wedge\cdots\wedge h^{i_{p-1}}\wedge
\bar h^{j_1}\wedge\cdots\wedge \bar h^{j_q}\,,\label{pqans2}
\eea   
where
\be
h^i \equiv \sigma_i\, \cosh r + \im\, \td\sigma_i\, \sinh r\,.\label{hdef}
\ee
It is easy also to verify that
\be
dh^i = \ft12 (\tanh r + \coth r)\, (dr - \im\, \nu)\wedge h^i + \ft12
   (\tanh r - \coth r)\, (dr-\im\, \nu)\wedge \bar h^i\,.
\ee

    Imposing $dG_{\sst{(p,q)}}=0$, we now find that the functions
$f_1$ and $f_2$ satisfy the equations
\bea
&& f_1' + f_2' + 2 (p\, f_1 + q\, f_2)\, \coth r =0\nn\,,\\
&& f_1' - f_2' + 2 (p\, f_1 - q\, f_2)\, \tanh r =0\,.
\eea
These equations can be solved in terms of hypergeometric functions, to
give
\bea
f_1 &=& c_1\, q\, _2F_1\left[ \ft12 p, \ft12 (q+1), \ft12 (p+q) +1;
-(\sinh 2r)^2\right]\nn\\
&& + c_2 \, (\sinh 2r)^{-p-q}\, 
_2F_1\left[ \ft12(1-p), -\ft12 q, 1-\ft12(p+q); -(\sinh
2r)^2\right]\,,\nn\\
f_2 &=& -c_1\, p\, _2F_1\left[ \ft12 q, \ft12 (p+1), \ft12 (p+q) +1;
-(\sinh 2r)^2\right] \nn\\
&&+ c_2 \, (\sinh 2r)^{-p-q}\, 
_2F_1\left[ \ft12(1-q), -\ft12 p, 1-\ft12(p+q); -(\sinh
2r)^2\right]\,,\label{pqsol}
\eea
where $c_1$ and $c_2$ are arbitrary constants.  Note that for any
specific choice of the integers $p$ and $q$ these expressions reduce
to elementary functions of $r$, so the occurrence of hypergeometric
functions here is just an artefact of writing formulae valid for all
$p$ and $q$.

\subsection{$L^2$-normalisable harmonic $(p,p)$-forms in $4p$ dimensions}

    In the special case where $p=q$, the above construction gives an
harmonic $(p,p)$-form in the middle dimension of a Stenzel manifold of
dimension $4p$.  In this case, we find that with $c_2$ taken to be zero,
the functions $f_1$ and $f_2$ in (\ref{pqsol}) become
\be
f_1 = -f_2= \fft{p\, c_1}{(\cosh r)^{2p}} \,,
\ee
and so the harmonic $(p,p)$-form $G_{\sst{(p,p)}}$ is given by
\bea
G_{\sst{(p,p)}} &=& \fft1{(\cosh r)^{2p}}\,
\ep_{i_1\cdots i_{p-1}j_1\cdots j_{p}}\,
\bar \ep^0\wedge \bar \ep^{i_1}\wedge\cdots \wedge\bar\ep^{i_{p-1}}\wedge
\ep^{j_1}\wedge\cdots\wedge\ep^{j_{p}}\nn\\
&& -\fft1{(\cosh r)^{2p}}\,   \,\ep_{i_1\cdots i_{p-1}j_1\cdots j_{p}}\,
\ep^0\wedge \ep^{i_1}\wedge\cdots\wedge \ep^{i_{p-1}}\wedge
\bar \ep^{j_1}\wedge\cdots\wedge \bar \ep^{j_q}\,,\label{ppharm}
\eea   
(after scaling out an irrelevant constant factor.)  It therefore has
magnitude given by
\be
|G_{\sst{(p,p)}}|^2 = \fft{\hbox{constant}}{(\cosh r)^{4p}}\,.
\ee
Since the $2(n+1)$-dimensional Stenzel metric has $\sqrt{g} =
\ft1{n+1}\, (\sinh 2r)^n$, and $n=2p-1$ here, it follows that this
harmonic form is $L^2$-normalisable (see footnote 2).

   One can also express this normalisable harmonic form in terms of
the original real vielbein basis.  Doing so, we find
\crampest
\bea
G_{\sst{(p,p)}}&=& \fft1{(\cosh r)^{n+1}}\, \times \nn\\
&&\sum_{s=0}^m \fft{m!}{s!\, (m-s)!}\, \Big[  \ep_{i_1\cdots i_{2s+1}
j_1\cdots j_{2m-2s}}\, e^{\td 0} \wedge e^{i_1}\wedge \cdots \wedge 
e^{i_{2s+1}}\wedge e^{\td j_1}\wedge\dots \wedge e^{\td j_{2m-2s}}
\nn\\
&&\qquad+  \ep_{i_1\cdots i_{2m -2s}
j_1\cdots j_{2s+1}}\, e^{0} \wedge e^{i_1}\wedge \cdots \wedge 
e^{i_{2m-2s}}\wedge e^{\td j_1}\wedge c\dots \wedge e^{\td j_{2s+1}}
\Big]\,,\label{l2harm}
\eea
\uncramp
where $m$ is defined by $n=2m+1=2p-1$.  In fact another way to obtain
the middle-dimension harmonic form when $n$ is odd is to write down an
ansatz of the form (\ref{l2harm}), with a different function of $r$
for each term in the sum, and then impose closure.  This leads to a
recursive system of first-order differential equations for the
functions, whose only solution giving an $L^2$ harmonic form is
(\ref{l2harm}).

\subsection{Non-normalisable harmonic $(p,q)$-forms}

   We saw above that the special case of a harmonic $(p,p)$-form in a
Stenzel manifold of dimension $4p$ yields the simple expression
(\ref{ppharm}) for an $L^2$-normalisable form.  It is not hard to see
that for any case other than $p=q$, the construction in section 3.1
always gives harmonic $(p,q)$-forms that are not $L^2$-normalisable.  A
divergence in the integral of $|G_{\sst{(p,q)}}|^2$ at $r=0$ is avoided
if the constant $c_2$ in (\ref{pqsol}) is chosen to be zero, but the
integral diverges at large $r$ unless $p=q$ (which can only occur in
dimensions that are a multiple of 4, since in general the dimension is
$2(p+q)$).  In fact the degree of divergence becomes larger as $|p-q|$
becomes larger.

    It follows from the above discussion that the ``most nearly
normalisable'' harmonic $(p,q)$-form in a Stenzel manifold of dimension
$4N+2$ will be for the case $(p,q)=(N+1,N)$ (or its complex conjugate).
One then finds that with $c_2=0$ the term involving $f_2$ dominates at
large $r$, and that
\be
|G_{\sst{(N+1,N)}}|^2 \sim \fft1{(\sinh 2r)^{2N}}\,.
\ee
Since we have $\sqrt{g} = (\sinh 2r)^n/(n+1)$, and $n=2N$ here, it
follows that the harmonic $(N+1,N)$-form is marginally not $L^2$
normalisable, and the integral of $|G_{\sst{(2N+1)}}|^2$ diverges as the
logarithm of the proper distance, at large radius.

   Our findings for $(p,q)$ middle-dimension harmonic forms, and
especially, the fact that only in dimensions $4N$ can there exist
$L^2$ harmonic forms, are consistent with the general discussion in
section 2.1.

\subsection{Canonical form, and special Lagrangian submanifold}

    If we take $q=0$, implying that $p=n+1$ in the
$2(n+1)$-dimensional Stenzel manifold, then with $c_2=0$ we see from 
(\ref{pqsol}) that $f_1$ vanishes, while $f_2$ becomes a constant.
This give the so-called canonical form, of type $(n+1,0)$:
\be
G_{\sst{(n+1,0)}} = \ep^0\wedge \ep^1 \wedge \cdots\wedge \ep^n\,.
\ee
It is easily verified that this is covariantly constant.  From
(\ref{pqans2}) and (\ref{hdef}) we see that it restricts to
\be
 -\im\, \nu\wedge\sigma_1\wedge \cdots \wedge \sigma_n
\ee
on the $S^{n+1}$ bolt at $r=0$.  Thus $\Re(G_\4)$ restricted to the
bolt vanishes.  We have already seen that the K\"ahler form
vanishes on the bolt, and so it follows that the bolt is a 
Special Lagrangian Submanifold.  Hence it is a calibrated submanifold,
and volume-minimising in its homology class; in other words, it is a
supersymmetric cycle.

\section{Applications: resolved M2-branes and D3-branes}

       The sequence of Stenzel metrics begins with $n=1$, which is the
4-dimensional Eguchi-Hanson metric.  It admits a normalisable harmonic
self-dual 2-form.  It was shown in \cite{clpres} that this can be used
to smooth out the the singularities in the heterotic 5-brane and in the
dyonic string, including the singularity that is associated with the
negative tension contribution in the dyonic string.  The resolved
solutions are smooth and supersymmetric, and have well-defined ADM
masses. We refer the reader to \cite{clpres} for details.

        In this section, we review the construction of the deformed
fractional D3-brane of \cite{klebstra}, which uses the 6-dimensional
Stenzel metric.  We also construct a new resolved M2-brane
using the 8-dimensional Stenzel metric.  Both solutions are smooth and
supersymmetric.  The D3-brane does not have a well-defined ADM mass,
whilst the M2-brane does.

\subsection{Fractional D3-brane using the 6-dimensional Stenzel metric}
 
     The standard D3-brane can be deformed when the six-dimensional
transverse space admits a harmonic self-dual 3-form.  In the notation we
shall use here, the general solution is given by \cite{clpres}
\bea
d\hat s_{10}^2 &=& H^{-1/2}\, dx^\mu\, dx^\nu\, \eta_{\mu\nu} +
H^{1/2}\, ds_6^2\,,\nn\\
F_\5 &=& d^4x\wedge dH^{-1} + \hat * dH\, \qquad
F_\3 = F_\3^{\rm RR} + \im\, F_\3^{\rm NS} = m\, G_\3\,,
\eea
where $ds_6^2$ is any six-dimensional Ricci-flat K\"ahler metric that
admits a non-trivial complex harmonic self-dual 3-form ${*G_\3} = \im \,
G_\3$, and $\hat *$ and $*$ are Hodge duals with respect to $d\hat
s_{10}^2$ and $ds_6^2$ respectively.  The function $H$ satisfies that
\be
\square H = -\ft{1}{12} m^2\, |G_\3|^2\,,
\ee
where $\square$ is the scalar Laplacian in the 6-dimensional transverse
space. 

    In \cite{klebstra}, a particular fractional D3-brane was
constructed where the six-dimensional Stenzel metric was used for the
transverse $ds_6^2$, and we shall now review this solution.  After
making trivial redefinitions (including $r\longrightarrow r/2$) in
order to adjust the conventions to those of \cite{klebstra}, and
taking $n=2$, the solution found in section 2.5 for the Stenzel metric
becomes
\bea
&&h^2 = \fft1{3K^2}\,,\qquad a^2 = 2K\, \cosh^2(r/2)\,,\qquad
b^2 = 2K\, \sinh^2(r/2)\,,\qquad c^2 = \fft{4}{3K^2}\,,\nn\\
&&\hbox{where}\qquad K = \fft{(\sinh 2r - 2r)^{1/3}}{2^{1/3}\, \sinh
r}\,,\label{d6met}
\eea
and the metric is then given by (\ref{stenmeth}) with $i$ running
over 2 values.  The Stenzel manifold is smooth, complete and
non-compact, with $r$ running from $r=0$ to $r=\infty$.  

   In these conventions, the general result (\ref{pqsol}) yields 
a harmonic $(2,1)$ form 
\be 
G_{\sst{(2,1)}} =\fft{2(r\, \coth r-1)}{\sinh^2 r}\,
\bar\ep^0\wedge\ep^1\wedge \ep^2 - \fft{(\sinh 2r -2r)}{2\sinh^3 r}\,
\ep^0\wedge (\ep^1\wedge \bar \ep^2 + \bar \ep^1\wedge
\ep^2)\,,\label{selfdual3form} 
\ee
This can be recognised as the self-dual harmonic 3-form constructed in
\cite{klebstra}, by noting that it can be expressed as
\be
G_\3 = \omega_\3 - \im\, {*\omega_\3}\,,
\ee
where
\be
\omega_\3 =  g_1\, e^0\wedge e^1 \wedge e^2 + g_2 \, 
   e^0\wedge e^{\td 1} \wedge e^{\td 2} + g_3 \,e^{\td 0} \wedge
   (e^1\wedge e^{\td 2} - e^2\wedge e^{\td 1})\,,
\ee
and
\be
g_1 =  \fft{\sinh r -r}{\sinh r\, \sinh^2(r/2)} \,,\quad 
g_2 =  \fft{\sinh r +r}{\sinh r\, \cosh^2(r/2)} \,,\quad 
g_3 = \fft{2(r\, \coth r -1)}{\sinh^2 r}\,.\label{fexp}
\ee

    Calculating the norm of $G_\3$, one obtains the result
\bea
|G_\3|^2 &=& 12g_1^2 + 12g_2^2 + 24 g_3^2\\
&=& \fft{96}{\sinh^6 r}\Big( (3+2\sinh^2 r)\, 
r^2 -3 r\,\sinh 2r +  3\sinh^2 r + \sinh^4 r  \Big)\,.\nn
\eea
Since for the metric we have $\sqrt{g} = \ft23 \sinh^2 r$, it follows
that $G_\3$ is not $L^2$ normalisable; it does not fall off
sufficiently rapidly at large $r$.

   It was argued in \cite{klebstra} that the self-dual harmonic 3-form
was of type $(2,1)$, and then in \cite{ganpol,gub} arguments were
presented that would show that the deformed D3-brane solution built
using $G_\3$ would be supersymmetric.  Our explicit proof that $G_\3$ is
of type $(2,1)$ thus demonstrates the supersymmetry of the solution.

     Because the $L^2$ norm of $G_\3$ converges for small $r$,
but diverges for large $r$, it follows that the function $H$ is
regular at small $r$, but does not fall off fast enough at large $r$ to
have a well-defined ADM mass.  In fact, $H$ has the large-$r$ asymptotic
behaviour given in (\ref{d3asymp}).

\subsection{Fractional M2-brane using the 
8-dimensional Stenzel metric}

    As a consequence of the Chern-Simons modification to the equation of
the motion of the 3-form potential in $D=11$ supergravity, namely
\be
d\hat {* F_\4} = \ft12 F_\4\wedge F_\4\,.
\ee
it is possible to construct a deformed M2-brane, given by
\cite{hawtay,deklm,clpres}
\bea 
d\hat s_{11}^2 &=& H^{-2/3}\, dx^\mu\, dx^\nu\, \eta_{\mu\nu} +
H^{1/3}\, ds_8^2\,,\nn\\
 F_\4 &=& d^3x\wedge dH^{-1} + m\, G_\4\,, 
\eea 
where $G_\4$ is the harmonic self-dual 4-form in the Ricci-flat
transverse space $ds_8^2$, and the function $H$ satisfies
\bea
\square H = -\ft1{48} m^2\, G_\4^2\,.
\eea
Warped reductions of this type, were also discussed in
\cite{gss,2beckers,becker}.

    In this section, we shall construct a deformed M2-brane using
the 8-dimensional Stenzel metric for the transverse $ds_8^2$.  In
this case, the index $i$ on $\sigma_i$ and $\td\sigma_i$ in the metric
(\ref{stenmeth}) runs over 3 values.  The Ricci-flat solution coming
from the first-order equations (\ref{firstorder}) is given by
\bea
&&a^2= \ft13 (2+\cosh 2r)^{1/4}\, \cosh r\,,\qquad 
b^2 =  \ft13 (2+\cosh 2r)^{1/4}\, \sinh r\, \tanh r\,,\nn\\
&&h^2=c^2 =  (2+\cosh 2r)^{-3/4}\, \cosh^3 r\,,\label{d8met}
\eea
with the metric then given by (\ref{stenmeth}).  The radial coordinate
runs from $r=0$ to $r=\infty$, and the metric lives on a smooth
complete non-compact manifold.

      In terms of the vielbein basis (\ref{stenviel}), we find from
(\ref{l2harm}) that the following is an $L^2$-normalisable self-dual
harmonic 4-form (of type $(2,2)$):
\bea
G_\4 &=& \fft3{\cosh^4 r}\, [e^{\td 0 }\wedge e^1\wedge e^2\wedge e^3
+ e^0\wedge e^{\td 1}\wedge e^{\td 2} \wedge e^{\td 3}] \nn\\
&&+ \fft1{2\cosh^4 r}\, \ep_{ijk} \, [e^0 \wedge e^i\wedge e^j \wedge
e^{\td k} + e^{\td 0}\wedge e^i \wedge e^{\td j} \wedge e^{\td k}]\,.
\label{harm4}
\eea
We can easily see that
\be
|G_\4|^2 = \fft{360}{\cosh^8 r}\,.
\ee

     The 8-dimensional Stenzel manifold can be used as the transverse
space to construct the deformed M2-brane.  The solution is given by
\bea
ds_{11}^2 &=& H^{-2/3}\, (-dt^2 + dx_1^2 + dx_2^2) + H^{1/3}\,
ds_8^2\,,\nn\\
F_\4 &=& dt\wedge dx_1\wedge dx_2\wedge dH^{-1} + m\, 
G_\4\,.\label{m2sol}
\eea
All the equations of motions are satisfied provided that
\be
\square H = -\ft1{48} m^2\, |G_\4|^2\,,\label{heqn}
\ee
where $\square$ is the scalar Laplacian in the 8-dimensional transverse
space.  Since we have $\sqrt{g} = \ft1{216}\sinh^3 (2r)$, assuming that
$H$ depends only on $r$, we have
\be
(h^{-2}\, \sqrt{g}\, H')' = -\fft{5m^2\, (\sinh 2r)^3}{144\cosh^8r}\,.
\ee
The first integration can be performed straightforwardly, giving
\be
H' = \fft{h^2}{\sqrt g}
\Big(\beta + \fft{7m^2\, \cosh 2r}{72 \cosh^4r}\Big)\,,
\ee
where $\beta$ is an arbitrary integration constant.  In order for the
solution regular at $r=0$, we must have
\be
\beta=-\fft{5m^2}{72}\,.\label{betavalue}
\ee
It is easier to perform the next integration by making a coordinate
redefinition,
\be
2+ \cosh 2r = y^4\,.
\ee
In terms of $y$, with $\beta$ given in (\ref{betavalue}),
the function $H$ is then given by
\bea
H&=&  -\fft{15 m^2}{\sqrt2}\, \int \fft{dy}{(y^4-1)^{5/2}}\nn\\
&=&c_0- \fft{5m^2\,(5y^5 - 7y)}{4\sqrt2 (y^4-1)^{3/2}} + 
\fft{25m^2}{4\sqrt2} F(\arcsin(\fft1{y})|-1)\,.
\eea
where $c_0$ is an integration constant, and $F(\phi|m)$ is the
incomplete elliptic integral of the first kind,
\be
F(\phi|m) \equiv \int_0^\phi (1- m\, \sin^2\theta)^{-1/2}\, d\theta\,.
\ee
It is easy to verify that the function $H$ is regular for $r$ running
from 0 to infinity.  For
$r=0$, $H$ is just a constant.  At large $r$, the function $H$ behaves
as
\be
H = c_0 + \fft{640m^2}{2187\rho^6} -
\fft{20480\, 2^{1/3}\, m^2}{28431\, 3^{2/3}\, \rho^{26/3}} +
\cdots\,,
\ee
where $\rho$ is the proper distance, defined by $h\, dr = d\rho$.  Thus
the M2-brane has no singularity, and it has a well-defined ADM mass.

     It is worth commenting further on the choice (\ref{betavalue}) for
the integration constant $\beta$.  The solution to (\ref{heqn})
has two integration constants $\beta$ and $c_0$, which originate from
the fact that one can add to $H$ any solution $H_0$ of the homogeneous 
equation
\be
\square H_0 =0\,.
\ee
However, the solution for $H_0$ has a singularity at small distance,
and so it requires an external delta-function source at the
singularity.  For the usual M2-brane with flat transverse space, the
divergence of $H$ signals a breakdown of our coordinate system, and
the delta function corresponds to a smooth horizon.  However, if, as
in the present case, the transverse space is a smooth non-flat
manifold, the singularity is real, and not just a coordinate artefact.
In the case when there is a $|G_\4|^2$ source, it is possible to find a
smooth everywhere-bounded positive function $H$.  In this case there
is no breakdown of our coordinate system, nor is there a naked
singularity, and we get a complete non-singular solution without an
horizon.\footnote{If the transverse space is a cone, and the harmonic
function depends only on the radial coordinate, the singularity in
$H_0$ corresponds to AdS$_4$ times the base of the cone.}  The $L^2$
normalisability then guarantees finiteness of the ADM mass, as may be
easily seen by integrating (\ref{heqn}).  Thus our choice for the
constant $\beta$ in (\ref{betavalue}) ensures that our solution is not
only smooth and non-singular, but it is also free of any horizon, and
is a rigorous supergravity solution.  Any other choice of the constant
$\beta$ would give a solution that was singular, requiring an external
source at the spacetime singularity on the horizon.

   Let us now consider the supersymmetry of the deformed solution.
From the $D=11$ supersymmetry transformations, it follows that if any
supersymmetry is to be preserved, the harmonic 4-form must satisfy:
\be
\delta\psi_a = \ft1{288} \Big( G_{bcde}\, \Gamma_{abcde} - 8
G_{abcd}\, \Gamma_{bcd} \Big) \, \eta =0\,.
\ee
Multiplying by $\Gamma^a$, we deduce that the two terms separately
must give zero, and in fact the supersymmetry condition can be reduced
to \cite{hawtay,2beckers}
\be
G_{abcd}\, \Gamma_{bcd}\, \eta=0\,.\label{d8susycon}
\ee
Now from (\ref{harm4}), the vielbein components of the 4-form are
given by
\be
G_{\td 0 ijk} = 3 u\, \ep_{ijk}\,,\quad
G_{0\td i\td j\td k} = 3u\, \ep_{ijk}\,,\quad
G_{0 ij\td k} = u\, \ep_{ijk}\,,\quad G_{\td0  i \td j\td k} = u\,
\ep_{ijk}\,,
\ee
where $u\equiv 1/\cosh^4 r$. Substituting into (\ref{d8susycon}), we
see that taking $a=0$, $i$, $\td i$ and $\td 0$ respectively, we
obtain the following conditions that must be satisfied if there is to
be preserved supersymmetry:
\bea
a=0:&& \ep_{ijk}\, (\Gamma_{\td i\td j\td k} + \Gamma_{ij\td k})\,
\eta=0
\,,\nn\\
a=i:&& \ep_{ijk}\, (3\Gamma_{\td 0 jk} + 2 \Gamma_{0 j\td k} +
\Gamma_{\td 0 \td j\td k})\, \eta=0\,,\nn\\
a=\td i:&& \ep_{ijk}\, (3\Gamma_{0 \td j\td k} + 2 \Gamma_{\td 0 \td j
k} + \Gamma_{0 j k})\, \eta=0\,,\nn\\
a=\td 0:&& \ep_{ijk}\, (\Gamma_{ijk} + \Gamma_{i\td j\td k})\,
\eta=0\,.\label{d8susy}
\eea
It is now a simple matter to show, using the integrability conditions
(\ref{gammacon}) which we already established, that the equations
(\ref{d8susy}) are satisfied, for both of the covariantly-constant
spinors on the Stenzel 8-manifold.  In other words, turning on the
deforming flux from the harmonic 4-form $G_\4$ does not lead to any
further breaking of supersymmetry, and so the resolved 
M2-brane preserves $\ft14$ of the original supersymmetry.

\section{Ricci-flat K\"ahler metrics on $\C^k$ bundles}

    There are many possible ans\"atze that one can adopt for
constructing classes of Ricci-flat metrics.  A classic procedure is to
look for metrics of cohomogeneity one, in which there are level
surfaces composed of homogeneous manifolds, with arbitrary functions
of radius parameterising homogeneous deformations of these
surfaces.\footnote{See \cite{danwan,danwan2} for a general discussion of such
metrics.}  The conditions for Ricci-flatness then reduce to ordinary
second-order differential equations for these functions.  If one is
lucky, the equations are solvable and the solutions include ones that
describe metrics on smooth complete manifolds.  Indeed, the Stenzel
construction that we studied in section 2 is an example of this type.
In cases where there are Ricci-flat solutions with special holonomy,
such as hyper-K\"ahler, K\"ahler or the $G_2$ and Spin(7) exceptional
cases, we have always found that first-order equations, derivable from a
superpotential, can be constructed.  All solutions of these satisfy the
second-order equations, but the converse is not necessarily true.

    In this section we study another general class of metrics of
cohomogeneity one, where the level surfaces are taken to be $U(1)$
bundles over a product of $N$ Einstein-K\"ahler manifolds, which would
typically themselves be homogeneous.  We then introduce $(N+1)$
arbitrary functions of the radial coordinate $r$, parameterising the
volumes of the $N$ base-space factors, and the length of the $U(1)$
fibres.  Following the familiar pattern, we then calculate the
curvature, derive the second-order equations for Ricci-flatness, and
then look for a first-order system coming from a superpotential.
Having done this, we are able to solve the equations and obtain
complete non-compact Ricci-flat K\"ahler metrics.  

   The Ricci-flat solutions that we obtain here 
are such that the metric coefficient for 
one of the factors in the base space goes to zero at $r=0$, as
does the coefficient in the $U(1)$ fibre direction.  This implies that
this particular factor in the base space must be a complex projective
space $\CP^m$, so that $r=0$ can become the origin of spherical polar
coordinates on $\R^{2k}$, where $k=m+1$.  If we write the base space
as ${\cal M} = \CP^m\times \wtd{\cal M}$, where $\wtd{\cal M}$ denotes
the product of the remaining Einstein-K\"ahler
manifolds in the base, then the total manifold has the topology
of a $\C^k$ bundle over $\wtd{\cal M}$.  The manifold has a bolt with
the topology $\wtd{\cal M}$ at $r=0$.   Global considerations impose
constraints on the possible choices for the other Einstein-K\"ahler
base space factors.  See \cite{d2frac} for a detailed discussion.

   Our principal focus will be on the case where all the
Einstein-K\"ahler factors in the base space are taken to be complex
projective spaces $\CP^{m_i}$.  It is shown in \cite{d2frac} that if
the metric coefficient for the $\CP^{m_1}$ factor is the one that goes
to zero at $r=0$, then regularity at $r=0$ implies that the other
$\CP^{m_i}$ factors must be such that
\be
m_1+1 = \hbox{gcd}(m_1+1,m_2+1,m_3+1,\ldots)\,,\label{gcdcon}
\ee
where $\hbox{gcd}$ denotes the greatest common divisor of its arguments.
The special case of just two factors, with the first being the trivial
zero-dimensional manifold $\CP^0$, and $\wtd{\cal M}$ being
$\CP^{m}$, gives a well-known sequence of Ricci-flat manifolds on
the complex line bundle over $\CP^{m}$.  The $m=1$ case is the
Eguchi-Hanson instanton.  We obtain an $L^2$-normalisable harmonic
$(m+1)$-form for all the $\C^{m+1}$ bundles over $\CP^m$ where $m$ is odd.

   The special case of two factors $\CP^{m_1}\times \CP^{m_2}$ with
$m_1=m_2=1$, for which the base space is $S^2\times S^2$ and the
topology of the total space is a $\C^2$ bundle over $\CP^1$, is the
6-dimensional ``small resolution'' of the conifold discussed in \cite{candel},
and more recently in \cite{zaytse}, as an alternative to the
``deformation'' of the conifold.  We shall study this in some detail,
and show that the non-normalisable harmonic 3-form used in
\cite{zaytse} to construct a fractional D3-brane gives a
non-supersymmetric solution.  We shall also consider three other
special cases in some detail, giving 8-dimensional examples where the
base space is $S^2\times \CP^2$ or 
$S^2\times S^2\times S^2$.\footnote{Note that the choice $S^2\times
\CP^2$ for the base actually violates the condition (\ref{gcdcon}) for
regularity when one of the factors collapses at the origin.  This
means that the metric actually has orbifold-like singularities at
$r=0$.  We shall discuss a way to avoid this difficulty later.}
construct $L^2$-normalisable harmonic 4-forms in two of these
manifolds, and use them to build further supersymmetric
deformed M2-branes.

\subsection{Curvature calculations, and superpotential}

    To begin with, since it illustrates most of the key features, we
shall consider the case of a base space that is the product of just two
factors, comprising Einstein-K\"ahler spaces of real dimensions $n$ and
$\td n$.  In the next subsection, we shall present the general results
for a product of $N$ Einstein-K\"ahler spaces.

   We make the following ansatz for metrics of cohomogeneity one
whose level surfaces are $U(1)$ bundles over 
products of two Einstein-K\"ahler base spaces:
\be
d\hat s^2 = dt^2 + a^2\, ds^2 + b^2\, d\td s^2 + c^2\, \sigma^2,
\ee
where $a$, $b$ and $c$ are functions of the radial coordinate $t$,
$ds^2$ and $d\td s^2$ are Einstein-K\"ahler spaces of real dimensions
$n$ and $\td n$ respectively, and
\be
\sigma = d\psi + A +\wtd A\,.
\ee
The potentials $A$ and $\wtd A$, living in $ds^2$ and $d\td s^2$
respectively, have field strengths $F=dA$ and $\wtd F=d\wtd A$,
given by $F= p\, J$, $\wtd F = q\, \wtd J$, where $J$ and $\wtd J$ are
the K\"ahler forms on $ds^2$ and $d\td s^2$.  Furthermore, we assume
cosmological constants $\lambda$ and $\td\lambda$ for the two spaces, so
\be
R_{ij} = \lambda\, \delta_{ij}\,,\qquad \wtd R_{ab} = \td\lambda\,
\delta_{ab}\,, \qquad
F_{ik}\, F_{jk} = p^2\, \delta_{ij}\,,\qquad \wtd F_{ac}\, \wtd
F_{bc} = q^2\, \delta_{ab}\,.
\ee
Note that there is a considerable redundancy in the use of constants
here, since $\lambda$ and $\td\lambda$ could be absorbed into rescalings
of the functions $a$ and $b$.  It is advantageous to keep all the
constants $\lambda$, $\td\lambda$, $p$ and $q$ unfixed for now, since
the choice of how to specify them most conveniently depends on what
choice one makes for the Einstein-K\"ahler metrics in the base space.

   In the orthonormal basis
\be
\hat e^0 = dt\,,\qquad \hat e^{\td 0} = c\, \sigma\,,\qquad 
\hat e^i = a\, e^i\,,\qquad \hat e^a = b\, e^a\,,\label{ob1}
\ee
we find that the non-vanishing components of the Ricci tensor are
\bea
\hat R_{00} &=& -n\, \fft{\ddot a}{a} -\td n\, \fft{\ddot b}{b} 
               -\fft{\ddot c}{c} \,,\nn\\
\hat R_{\td 0 \td 0} &=& -n\, \fft{\dot a\, \dot c}{a \, c} -\td n \,
\fft{\dot b\, \dot c}{b\, c} - \fft{\ddot c}{c} + \fft{n\, p^2\,
c^2}{4 a^4} + \fft{\td n\, q^2\, c^2}{4 b^4} \,,\label{tzgenric}\\
\hat R_{ij} &=& -\Big( \fft{\ddot a}{a} + \fft{\dot a\, \dot c}{a\, c} 
+ (n-1)
\, \fft{{\dot a}^2}{a^2} + \td n\, \fft{\dot a\, \dot b}{a\, b} - 
\fft{\lambda}{a^2} + \fft{p^2\, c^2}{2 a^4}\Big)\, \delta_{ij}\,,\nn\\
\hat R_{ab} &=& -\Big( \fft{\ddot b}{b} + \fft{\dot b\, \dot c}{b\, c} +
(\td n-1) \, \fft{{\dot b}^2}{b^2} + n\, \fft{\dot a\, \dot b}{a\, b} - 
\fft{\td\lambda}{b^2} + \fft{q^2\, c^2}{2 b^4}\Big)\, \delta_{ab}\,.\nn
\eea
From this, after introducing the new radial variable $\eta$ defined
by $dt= c\, a^n\, b^{\td n}\, d\eta$, we find that the conditions for
Ricci-flatness can be derived from the Lagrangian $L=T-V$, where
\bea
T &=& n\,  \a'\, \gamma' + \td n \, \beta'\, \gamma' + n\, \td n\, \a'\,
\beta' + \ft12 n(n-1)\, {\a'}^2 + \ft12\td n(\td n-1)\, {\beta'}^2 
 \,,\nn\\
V &=& \ft18 n\, p^2\, e^{(2n-4)\, \a + 2\td n\, \beta + 4\gamma} +
   \ft18 \td n\, q^2\, e^{2n\, \a + (2\td n-4) \, \beta + 4\gamma}
\nn\\
&&- \ft12 n\, \lambda\, e^{(2n-2)\, \a + 2\td n\, \beta + 2\gamma} -
 \ft12 \td n\, \td \lambda\, e^{2n\, \a + 
(2\td n-2)\, \beta + 2\gamma} \,,
\eea
together with the requirement that $T+V$ vanishes.  Here,
a prime means a derivative with respect to $\eta$.

   Defining $\a^i= (\a, \beta, \gamma)$ as usual, we find that the
Lagrangian can be written as $L=\ft12 g_{ij} \, (d{\a^i}/d\eta)\, 
(d{\a^j}/d\eta) + \ft12 g^{ij}\, \del W/\del \a^i\, \del W/\del\a^j$, where the
superpotential is given by
\be
W = \ft14 n\, p\, e^{(n-2)\, \a + \td n \, \beta + 2\gamma} +
   \ft14 \td n\, q\, e^{n\, \a + (\td n-2)  \, \beta + 2\gamma} +
   k\, e^{n\, \a + \td n\, \beta}
\ee
and the various constants must be chosen so that
\be
k = \fft{\lambda}{p} = \fft{\td\lambda}{q}\,.\label{krel}
\ee
This leads to the first-order equations
\bea
&&\a' = \ft12 p\, e^{(n-2)\, \a\ + \td n\, \beta + 2\gamma}\,,\qquad
\beta' = \ft12 q\, e^{n\, \a + (\td n-2)\, \beta +
2\gamma}\,,\label{tzgenfirst} \\
&&\gamma' = -\ft14 n\, p\, e^{(n-2)\, \a + \td n\, \beta + 2\gamma} -
\ft14 \td n\, q\, e^{n\, \a + (\td n-2)\, \beta + 2\gamma} + k\,
e^{n\, \a + \td n\, \beta}\,.\nn
\eea

\subsection{Solving the first-order equations}

   We proceed here by introducing a new radial variable $r$, defined
by\footnote{Note that another choice is to take $dr=e^{n\, \a + (\td
n-1)\, \beta + 2\gamma}\, d\eta$; this will reverse the r\^oles of the
two metrics $ds^2$ and $d\td s^2$, with consequences that will become
clear later.}
\be
dr = e^{(n-1)\, \a + \td n\, \beta + 2\gamma}\, d\eta\,.
\ee
The first-order equations (\ref{tzgenfirst}) now become
\be
\fft{d\a}{dr} = \ft12 p\, e^{-\a}\,,\quad
\fft{d\beta}{dr} = \ft12 q\, e^{\a-2\beta}\,,\quad
\fft{d\gamma}{dr} = -\ft14 n\, p\, e^{-\a} - \ft14 \td n\, q\,
e^{\a-2\beta} + k\, e^{\a-2\gamma}\,.\nn
\ee
The first can be solved at sight; the second can then be solved, and
then using these results the third can be solved.  After making an
appropriate choice of integration constants, the result is
\bea
e^{2\a} &=& \ft14 p^2\, r^2\,,\qquad e^{2\beta} = 
\ft14 p\, q\, (r^2+ \ell^2)\,,\nn\\
e^{2\gamma} &=& \fft{k\, p\, r^2}{n+2}\, 
\Big(1+\fft{r^2}{\ell^2}\Big)^{-\td
n/2}\, _2F_1\Big[1+\ft12 n, -\ft12 \td n, 2+\ft12 n,
-\fft{r^2}{\ell^2}\Big]\,,\label{tzgensol}
\eea
where $\ell$ is a constant.  The Ricci-flat metric is given by
\be
d\hat s^2 = e^{2\a-2\gamma}\, dr^2 + e^{2\gamma}\, \sigma^2 +
e^{2\a}\, ds^2 + e^{2\beta}\, d\td s^2\,.\label{tzgenmet}
\ee
(Note that once one plugs in specific integer values for $n$ and $\td
n$, the hypergeometric function in the expression for $e^{2\gamma}$
becomes purely algebraic.)

   At small $r$, we have
\be
e^{2\a} = \ft14 p^2\, r^2\,,\qquad e^{2\beta} \sim \ft14 p\, q\,
\ell^2\,,\qquad e^{2\gamma}\sim \fft{k\, p}{n+2}\, r^2\,.
\ee
Bearing in mind that $k=\lambda/p$, we therefore find that near
$r=0$, the metric approaches
\be
d\hat s^2 \sim \fft{(n+2)\, p^2}{4\lambda}\, dS^2 + \ft14 p\, q\,
\ell^2\, d\td s^2\,,
\ee
where
\be
dS^2 = dr^2 + r^2\, \Big(\fft{4\lambda^2}{p^2\, (n+2)^2}\, \sigma^2 +
\fft{\lambda}{n+2}\, ds^2\Big)\,.
\ee
Regularity at $r=0$ therefore requires that the quantity enclosed in the
parentheses be the unit $(n+1)$-sphere metric. This means in particular
that $ds^2$ should be the standard Fubini-Study metric on $\CP^{m}$,
where $n=2m$.  The canonical choice for the cosmological constant that
gives a ``unit'' $\CP^{m}$ is in fact
\be
\lambda= n+2\,,
\ee
and the Fubini-Study metric is then $ds^2=d\Sigma_m^2$, where
\be
d\Sigma_m^2 = F^{-1}\, d\bar z^{a}\, dz^a -F^{-2}\, 
\bar z^a\, z^b\, dz^a\, d\bar z^b\,,\label{cpmmet}
\ee
and $F=1+\bar z^a\, z^a$.  After setting $\lambda=n+2$, we therefore
find that
\be
d\Omega^2 \equiv \fft{4}{p^2} \, \sigma^2 + ds^2\,,
\ee
should be the unit $(2m+1)$-sphere metric.  The $\wtd A$ term in
$\sigma$ is irrelevant here, and so regularity demands that
\be
d\Omega^2 = \Big( d\psi + \fft{2}{p}\, A\Big)^2 + ds^2
\ee
must be the unit $(n+1)$-sphere.  Recalling that we originally required
that $dA= p\, J$, where $J$ is the K\"ahler form on $ds^2$, we see that
this means that regularity requires that the potential $B$ in $d\Omega^2
= (d\psi + B)^2 + ds^2$ should give $dB=2J$.  This is precisely what one
finds in the description of $S^{2m+1}$ as the Hopf fibration over
$\CP^m$.   More detailed discussions  of the
regularity conditions at $r=0$ are discussed in \cite{d2frac}:  In order for the fibre coordinate to
have the correct periodicity, it must be that the other Einstein-K\"ahler
factor must impose no further restriction above those implied by the
$\CP^m$ itself.  For example, if the other factor is $\CP^{\td m}$,
then it must be that \cite{d2frac}
\be
m+1=\hbox{gcd}(m+1,\td m+1)\,.\label{cpcp}
\ee

    We can summarise the above results as follows.  We have found that
the Ricci-flat metric given by (\ref{tzgensol}) and (\ref{tzgenmet})
can be regular at $r=0$, if the $n$-dimensional Einstein-K\"ahler
metric $ds^2$ is taken to be the Fubini-Study metric on $\CP^m$, with
$n=2m$.  Furthermore the Einstein-K\"ahler manifold for the metric
$d\td s^2$ must satisfy certain topological conditions \cite{d2frac},
which reduce to (\ref{cpcp}) if it is $\CP^{\td m}$.  Then $r=0$ is a
regular region in the manifold, corresponding to a bolt whose topology
is that of the Einstein-K\"ahler manifold with metric $d\td s^2$.  For
$r>0$, we have level surfaces that are $U(1)$ bundles over the product
of the two Einstein-K\"ahler spaces whose metrics are $ds^2$ and $d\td
s^2$.

   Of course the constants $p$ and $q$ must be chosen appropriately, to
be commensurate with the periodicity of the fibre coordinate $\psi$.  For
example, if one takes the base space to be the product $\CP^{m}\times
\CP^{\td m}$, and chooses the canonical values $\lambda= 2(m+1)$ and
$\td\lambda = 2(\td m+1)$ for the cosmological constants so as to give
unit Fubini-Study metrics, then, after taking into account the relation
(\ref{krel}), we may without loss of generality take $p=m+1$, $q=\td
m+1$.  The fibre coordinate $z$ must then have period $2\pi$, implying
that the $U(1)$ bundle over $\CP^{m}\times \CP^{\td m}$ is
simply-connected, or $2\pi/s$, where $s$ is any integer, in which case
the bundle space is not simply connected.\footnote{ See, for example,
\cite{hoxmarpop} for a detailed discussion.  It is also shown in
\cite{hoxmarpop} that these specific $U(1)$ bundles over $\CP^{m}\times
\CP^{\td m}$ admit Killing spinors when the scalings are chosen so that
the metric is Einstein.}  Thus when we consider $\CP^{m}\times \CP^{\td
m}$ base spaces, we shall typically make the choices
\be 
\lambda= 2(m+1)\,,\quad \td\lambda = 2(\td m+1)\,,\quad p=
m+1\,,\qquad q=\td m+1\,.\label{cpmconv}
\ee

   We can, of course, consider instead the situation where the r\^oles
of the two metrics $ds^2$ and $d\td s^2$ are interchanged, as mentioned
in the footnote above.  Everything goes through, {\it mutatis mutandis},
in exactly the same way as described above.  It will now be the metric
$d\td s^2$ that is required to be the Fubini-Study metric on $\CP^{\td
m}$, with $\td n=2\td m$.

   Substituting the first-order equations (\ref{tzgenfirst}) back into 
the expressions for the curvature 2-forms, we can read off the
integrability conditions $\hat R_{ABCD}\, \Gamma_{CD}\, \eta=0$ for
the existence of covariantly-constant spinors.  These conditions give
\be
(\Gamma_{0i} + J_{ij}\, \Gamma_{\td 0 j})\, \eta=0\,,\qquad 
(\Gamma_{0a} + \wtd J_{ab}\, \Gamma_{\td 0 b})\, \eta=0\,.
\label{tzgenspin}
\ee
The spinors that satisfy these conditions are the expected complex pair
of covariantly-constant spinors in the Ricci-flat K\"ahler metrics.

   It is straightforward to establish that the K\"ahler form is given
by
\be
\hat J = \hat e^0\wedge\hat e^{\td 0} + e^{2\a}\, J + e^{2\beta}\,
\wtd J\,,\label{genkah}
\ee
or, in other words, the vielbein components $\hat J_{AB}$ are given by
$\hat J_{0\td 0} =1$, $\hat J_{ij} = J_{ij}$, $\hat J_{ab} = \wtd
J_{ab}$.  (As in our discussion of the Stenzel metrics, we could again
instead {\it derive} the first-order equations (\ref{tzgenfirst}) by
requiring that (\ref{genkah}) be covariantly constant.)

   It can be useful to obtain complex holomorphic coordinates $z^\mu$ for the 
K\"ahler metrics on the $\C^k$ bundle spaces.  This can be done by
solving the holomorphicity conditions $(\delta_M^N + 
\im\, J_M{}^N)\, \del_M\, z^\mu=0$.  It is most convenient to do this
in the orthonormal frame, for which we therefore need 
the basis of vector fields dual to the vielbein (\ref{ob1}).  Using
the $r$ coordinate of (\ref{tzgenmet}), it is given by
\bea
&&\hat E_0 = e^{\gamma-\a}\, \fft{\del}{\del r} \,,\qquad \hat E_{\td 0}
= e^{-\gamma}\, \fft{\del}{\del\psi}\,,\nn\\
&&\hat E_i= e^{-\a}\, \Big(E_i
- A_i\, \fft{\del}{\del \psi}\Big)\,,\qquad
\hat E_a= e^{-\beta}\, \Big(\wtd E_a
- \wtd A_a\, \fft{\del}{\del \psi}\Big)\,,
\eea
where $E_i$ and $\wtd E_a$ are the vielbeins inverse to $e^i$ and $\td
e^a$ on $ds^2$ and $d\td s^2$.  
From this we obtain the following holomorphicity conditions:
\bea
\Big(\fft{\del}{\del \rho} + \im\,\fft{\del}{\del\psi}\Big)\,
z^\mu &=& 0\,,\nn\\
(E_i + \im\, J_{ij}\, E_j)\, z^\mu - (A_i + \im\, J_{ij}\, A_j)\,
\fft{\del z^\mu}{\del\psi} &=&0\,,\label{holcon}\\
(\wtd E_a + \im\, \wtd J_{ab}\, \wtd E_b)\, z^\mu - 
(\wtd A_a + \im\, \wtd J_{ab}\, \wtd A_b)\,
\fft{\del z^\mu}{\del\psi} &=&0\,,\nn
\eea
where we have defined the new radial coordinate $\rho$ by
\be
d\rho = e^{\a-2\gamma}\, dr\,.
\ee
It is evident therefore that the holomorphic coordinates for the
K\"ahler metrics $ds^2$ and $d\td s^2$ themselves can be used as
holomorphic coordinates in the total manifold.  It therefore remains
to find one more complex coordinate $z^0$, constructed from the additional
real coordinates $\rho$ and $\psi$.  The first equation in
(\ref{holcon}) shows that $z^0$ should be a function of $\rho+\im\,
\psi$.  Noting that the vector potentials $A_i$ and $\wtd A_a$ can be
written in terms of the K\"ahler functions $K$ and $\wtd K$ on $ds^2$
and $d\td s^2$ as
\be
A_i = p\, J_i{}^j\, \del_j \, K\,,\qquad \wtd A_a= q\, \wtd J_a{}^b\, \del_b\,
\wtd K\,,
\ee
it therefore follows that the extra complex coordinate can be taken to
be
\be
z^0 = e^{\rho+\im\, \psi + p\, K + q\, \wtd K}\,.\label{z00sol}
\ee

   We conclude this subsection with a number of explicit examples.
Our first two make use of an $S^2\times \CP^2$ base space.  These
will actually still have orbifold-like singularities at $r=0$, as we
discussed above, since (\ref{cpcp}) is not satisfied.  However, they
are still of interest for
the purposes of constructing deformed M2-brane solutions, since the
remaining singularities are rather mild ones.  We shall discuss this,
and a complete resolution of the singularities, later.

\bigskip
\noindent\underline{{\bf $\C^2/\Z_2$ bundle over $\CP^2$}}:
\bigskip

   A particular class of examples would be to take the base space to
be $S^2\times \CP^2$, in which case we get 8-dimensional Ricci-flat
K\"ahler metrics.  Note that there are two distinct types of solution;
one of them has a $\CP^2$ bolt at $r=0$, whilst the other has instead
an $S^2$ bolt. 

    Consider first the case with the $\CP^2$ bolt; with our form of
the solution where the untilded metric is singled out as the one whose
coefficient goes to zero at $r=0$, we therefore take $ds^2$ to be the
$S^2$ metric, and $d\td s^2$ to be the $\CP^2$ metric.  From our
general results, after making the conventional choices
(\ref{cpmconv}), \ie $\lambda=4$, $\td\lambda=6$, $p=2$, $q=3$ here,
the Ricci-flat K\"ahler 8-metric is then given by (\ref{tzgenmet})
\be
e^{2\a} = r^2\,,\qquad e^{2\beta} = \ft32 (r^2+
\ell^2)\,,\qquad e^{2\gamma} = \fft{r^2\, (3 r^4 + 8 \ell^2\, r^2 + 6
\ell^4)}{ 6(r^2+\ell^2)^2}\,.
\ee
(Note that the unit $\CP^1$ is actually a 2-sphere of radius $\ft12$.)
Thus the Ricci-flat K\"ahler metric is
\be
d\hat s_8^2 = U^{-1}\, dr^2 +  r^2\, U \, \sigma^2 
+ \ft14 r^2\, (d\theta^2 +
\sin^2\theta\, d\phi^2) + \ft32 (r^2+ \ell^2)\, d\Sigma_2^2\,,
\label{s2cp2met}
\ee
where
\be
\sigma = d\psi - \ft12 \cos\theta\, d\phi + \wtd A\,,\qquad
U= \fft{3 r^4 + 8 \ell^2\, r^2 + 6 \ell^4}{ 6(r^2+\ell^2)^2} \,.
\ee
Here the maximum allowed periodicity for $\psi$ is 
$(\Delta\psi)_{\rm max}=\pi$ (see, for example, \cite{pagpopw}), 
and $d\wtd A= 3 \wtd J$, where $\wtd J$
is the K\"ahler form on the unit $\CP^2$ metric $d\Sigma_2^2$, given
in (\ref{cpmmet}).  If $\psi$ had had the period $2\pi$, then $U(1)$
fibres over $S^2$ would describe $S^3$, and the metric would approach
$\R^4\times \CP^2$ locally near $r=0$.  Instead, we get the lens space
$S^3/\Z_2$, and the metric therefore
approaches $(\R^4/\Z_2)\times \CP^2$; the 8-manifold is a $\C^2/\Z_2$
bundle over $\CP^2$.  We could, of course, replace $\CP^2$ by the
standard Einstein-K\"ahler metric on $S^2\times S^2$ in this metric,
in which case the fibre coordinate would be allowed to have the period
$2\pi$ needed for complete regularity at $r=0$.  In fact this would
give a special case of a more general class of Ricci-flat K\"ahler
metrics on $\C^2$ bundles over $S^2\times S^2$, which we shall
construct in section 5.3.

\bigskip
\noindent\underline{{\bf $\C^3/\Z_3$ bundle over $\CP^1$}}:
\bigskip

    The other possibility using $S^2\times \CP^2$ in the base space is
to interchange the r\^oles of the $S^2$ and $\CP^2$ factors, so that
now $ds^2$ is the $\CP^2$ metric, and $d\td s^2$ is the $S^2$ metric.
It is convenient to refer to this therefore as a $\CP^2\times S^2$
base, with the understanding that it is always the first factor whose
metric coefficient goes to zero at $r=0$.  For this example, it is
therefore convenient to choose the constants so that $\lambda=6$,
$\td\lambda=4$, $p=3$ and $q=2$.  The resulting Ricci-flat K\"ahler
8-metric is then
\be
d\hat s_8^2 = U^{-1}\, dr^2 + \ft94 r^2\, r^2\, U\, \sigma^2 + \ft94
r^2\, d\Sigma_2^2 + \ft32(r^2+\ell^2)\, d\Sigma_1^2\,,\label{cp2s2met}
\ee
where in this case we have
\be
\sigma = d\psi + A + \wtd A\,,
\qquad
U= \fft{3r^2+ 4\ell^2}{9(r^2+\ell^2)}\,,
\ee
and $dA=3J$, $d\wtd A=2\wtd J$.  The metrics $d\Sigma_2^2$ and
$d\Sigma_1^2$ are the unit metrics on $\CP^2$ and $\CP^1$
respectively, given by (\ref{cpmmet}), and $J$ and $\wtd J$ are their
respective K\"ahler forms.  (Note that $d\Sigma_1^2 = \ft14
d\Omega_2^2 = \ft14 (d\theta^2+\sin^2\theta\, d\phi^2)$.)  The maximum
allowed periodicity for $\psi$ is again $(\Delta\psi)_{\rm max} =
\pi$, while the period that would be need for the $U(1)$ bundle over
$\CP^2$ to describe $S^5$ would be $\Delta\psi=3\pi$.  This means that
we instead get the lens space $S^5/\Z_3$, and so near $r=0$ the metric
approaches $\R^6/Z_3\times S^2$; the 8-manifold is a $\C^3/\Z_3$
bundle over $S^2$ (or $\CP^1$).

\bigskip
\noindent\underline{{\bf Complex line  bundle over $\CP^m$}}:
\bigskip

   Another possibility is to take one factor in the product base
manifold to be trivial, and the other to be $\CP^m$ (or any other
Einstein-K\"ahler manifold).  The case where $m=1$ is Eguchi-Hanson;
for general $m$ the corresponding Ricci-flat K\"ahler metrics were
constructed in \cite{berber}, and also in \cite{pagpop}.  Since we
shall make use of one of these examples later, we shall summarise the
general results here.  By taking $p=n=\lambda=0$, $q=1$, $\td\lambda=
\td n+2=2m+2$, and setting $\a=0$, the first-order equations
(\ref{tzgenfirst}) can be solved to give the $2(m+1)$-dimensional
Ricci-flat K\"ahler metric
\be
d\hat s^2 = U^{-1}\, dr^2 + 4 r^2\, U\, \sigma^2 + r^2\,
d\Sigma_m^2\,,\label{bbmet}
\ee
where $r$ here is related to the $r$ variable in (\ref{tzgenfirst}) 
by $r\longrightarrow r^2$,
the function $U$ is given here by
\be
U= 1-\Big(\fft{r_0}{r}\Big)^{2m+2}\,,
\ee
with $r_0$ being a constant, and $d\Sigma_m^2$ is the metric on the
unit Fubini-Study metric on $\CP^m$, given in (\ref{cpmmet}).  Note
that $\sigma=d\psi+\wtd A$ here, where $d\wtd A=J$, the K\"ahler form on
the $\CP^m$.  The radial coordinate $r$ runs from $r=r_0$,
where the metric approaches $\R^2 \times \CP^m$, to infinity.
Topologically, the manifold is a $\C^1$ bundle over $\CP^m$.  

  For future reference we note that it is very easy to solve for an
$L^2$-normalisable (anti)-self-dual harmonic form in the middle
dimension, when $m$ is odd.  It is given by
\be
G_{\sst{(m+1)}} = \fft1{r^{2m+2}}\, \Big[ r^{m-1}\, 
\hat e^0\wedge \hat e^{\td 0} \wedge J^{(m-1)/2} - \fft{2}{m+1}\, 
 r^{m+1}\, J^{(m+1)/2} \Big]\,.\label{bbharm}
\ee
Note that the factors of $r$ within the square brackets just
convert each power of the K\"ahler form $J$ on $\CP^m$ into a 2-form
of unit magnitude in the metric $d\hat s^2$, \ie $r^2\, J = \ft12
J_{ab}\, \hat e^a\wedge \hat e^b$.  Thus each term within the square 
brackets is just a constant times a wedge product of hatted
vielbeins.  The magnitude of $G_{\sst{(m+1)}}$ is therefore given by
\be
|G_{\sst{(m+1)}}|^2 = \fft{\hbox{constant}}{r^{4m+4}}\,,
\label{bbharm2}
\ee
and so the $L^2$-normalisability is manifest.

\subsection{General results for $N$ Einstein-K\"ahler factors in the
base space}

   As we indicated above, the construction of the previous subsection 
has a straightforward generalisation to the case where we have $N$
Einstein-K\"ahler factors in the base space,
\be
{\cal M} = M_1\times M_2 \times \cdots \times M_N\,,\label{mprod}
\ee
with real dimensions $n_i$ and metrics $ds_i^2$.  Thus we write
\be
d\hat s^2 = dt^2 + \sum_{i=1}^N a_i^2\, ds_i^2 + c^2\, \sigma^2\,,
\ee
where 
\be
\sigma = d\psi + \sum_i A^i\,,
\ee
where $dA^i= p_i\, J^i$, and $J^i$ is the K\"ahler form on the factor
$M_i$ in the base manifold.  By comparing with the previous
subsection, our notation here and it what follows should be
self-evident.

   We find that the Ricci tensor for $d\hat s^2$ has components
\bea
\hat R_{00} &=& -\sum_i n_i\, \fft{\ddot a_i}{a}  
               -\fft{\ddot c}{c} \,,\nn\\
\hat R_{\td 0 \td 0} &=& -\sum_i n_i\, \fft{\dot a_i\, \dot c}{a_i \,
c} - \fft{\ddot c}{c} + \sum_i \fft{n_i\, p_i^2\,
c^2}{4 a_i^4} \,,\label{tzgenric2}\\
\hat R_{a_ib_i} &=& -\Big( \fft{\ddot a_i}{a_i} 
+ \fft{\dot a_i\, \dot c}{a_i\, c} -
 \fft{{\dot a_i}^2}{a_i^2} + \fft{\dot a_i}{a_i } \, \sum_j n_j
\fft{\dot a_j}{a_j}  -
\fft{\lambda_i}{a_i^2} + \fft{p_i^2\, c^2}{2 a_i^4}\Big)\, 
\delta_{a_ib_i}\,.\nn
\eea
Defining $a_i=e^{\a_i}$, $c=e^{\gamma}$, the conditions for
Ricci-flatness can be derived from the Lagrangian
\be
L = \ft12 \sum_{i,j} n_i\, n_j\, \a_i'\, \a_j' - 
\ft12 \sum_i n_i\, {\a_i'}^2 + \sum_i n_i\, \a_i'\, \gamma' -V\,,
\label{genlag}
\ee
where
\be
V= \ft18 \sum_i n_i\, p_i^2\, e^{2 \mu_i} -
\ft12 \sum_i n_i\, \lambda_i\, e^{2\mu_i + 2\a_i -2\gamma}\,,
\ee
with
\be
\mu_i \equiv 2\gamma - 2\a_i + \sum_j n_j\, \a_j\,.
\ee
The primes denote derivatives with respect to $\eta$, defined by
\be
dt = e^{\sum_i n_i\, \a_i + \gamma}\, d\eta\,.
\ee

   Defining $\a_0=\gamma$, and indices $a=(0,i)$, the Lagrangian
(\ref{genlag}) can be written as $L=\ft12 g_{ab} \, (d\a^a/d\eta) \,
(d\a^b/d\eta) -V$, with $g_{ij}= n_i\, n_j - n_i\, \delta_{ij}$,
$g_{0i}= n_i$, $g_{00}=0$. This has the inverse
\be
g^{ij} = \fft{1}{D}\,,\qquad g^{0i} = \fft{1}{D}\,,\qquad
g^{00} = \fft{1}{D} -1\,,
\ee
where $D=\sum_i n_i$ is the total dimension of the base space.
It is then straightforward to show that the potential $V$ can be
written in terms of a superpotential $W$, as $V=- \ft12 g^{ab}\, (\del
W/\del\a^a)\, (\del W/\del\a^b)$, where
\be
W = \ft1{4}\, \sum_i n_i\, p_i\, e^{\mu_i} + k\, 
e^{\sum_i n_i\, \a_i}\,,
\ee
provided that the constants $p_i$ and $\lambda_i$ satisfy
\be
\lambda_i = k\, p_i\,.\label{lamp}
\ee

    It follows that the following first-order equations imply
Ricci-flatness:
\be
\a_i' = \ft12 p_i\, e^{\mu_i}\,,\qquad \gamma' = k\, e^{\sum_i n_i\,
\a_i} -\ft14 \sum_i n_i\, p_i\, e^{\mu_i} \,.
\ee
We can solve these by defining a new radial coordinate\footnote{We
single out the $i=1$ factor in the base space purely as a matter of
convention; there is no loss of generality, since we have not 
yet specified the choices for these factors} $r$:
\be
dr = e^{\mu_1 + \a_1}\, d\eta\,,
\ee
which leads to
\be
\fft{d\a_i}{dr} = \ft12 p_i\, e^{\a_1- 2\a_i}\,,\qquad
\fft{d\gamma}{dr} = k\, e^{\a_1-2\gamma} -\ft14 \sum_i n_i\, p_i\,
e^{\a_1-2\a_i}\,.
\ee
The equation for $\a_1$ can be solved immediately, and then those for
the remaining $\a_i$ can be integrated.  We find
\be
e^{2\a_i} =  \ft14 p_1\, p_i\, (r^2+ \ell_i^2)\,,
\ee
where $\ell_1=0$ and the other $\ell_i$ are constants of integration.
Defining $\td\gamma\equiv \gamma+ \ft12\sum_i n_i\, \a_i$ in an
intermediate step, and $x\equiv r^2$, the equation for $\gamma$ can be
solved to give
\be
e^{2\gamma} = \ft12 p_1\, k\,\prod_i (x+\ell_i^2)^{-n_i/2}\, 
\int_0^x dy\, \prod_j (y+\ell_j^2)^{n_j/2}\,.
\ee
The integration is elementary, giving an expression for $e^{2\gamma}$
as a rational function of $x$ for any given choice of the integers
$n_i$, but the general expression for arbitrary dimensions $n_i$
requires the use of hypergeometric functions.  In terms of the $r$
coordinate, the metric is given by
\be
d\hat s^2 = e^{2\a_1-2\gamma}\, dr^2 + \sum_i e^{2\a_i}\, ds_i^2 + 
          e^{2\gamma}\, \sigma^2\,.
\ee

   The analysis of the structure of the Ricci-flat metrics proceeds in
a fashion that is analogous to that of the previous section.  The
radial coordinate runs from $r=0$, where the metric functions
$e^{2\a_1}$ and $e^{2\gamma}$ vanish, to $r=\infty$.  Regularity at
$r=0$ requires that the Einstein-K\"ahler metric $ds_1^2$ on the
factor $M_1$ in the base space (\ref{mprod}) be the Fubini-Study
metric on $\CP^{m_1}$, where $n_1=2m_1$, so that $r=0$ becomes the
origin of spherical polar coordinates on $\R^{n_1+2}$.  Even though
the other metric functions $e^{2\a_i}$ for $i\ge2$ are non-zero for
the entire range $0\le r\le\infty$, there are again topological
restrictions on the choice of Einstein-K\"ahler manifolds for these
factors, stemming from the requirement that the $U(1)$ fibre
coordinate should have the periodicity needed for the $U(1)$ bundle
over $\CP^{m_1}$ to be the $(2m_1+1)$-sphere and not a lens space.
(See \cite{d2frac} for a detailed discussion: If, for example, all the
other factors are complex projective spaces $\CP^{m_i}$, then they
must satisfy (\ref{gcdcon}).)  Topologically, the manifold on which
the metric $d\hat s^2$ is then defined is a $\C^k$ bundle over the
product of the remaining base-space factors $M_2\times M_3\times
\cdots \times M_N$, where $k=\ft12 n_1 + 1$.

   Arguments analogous to those of the previous subsection show tha
the K\"ahler form for the metric $d\hat s^2$ is given by
\be
\hat J = \hat e^0\wedge \hat e^{\td 0} + \sum_i a_i^2 \, J^i\,,
\ee
where $J^i$ denotes the K\"ahler form on the $i$'th factor in the
product of Einstein-K\"ahler manifolds (\ref{mprod}) in the base space.
The two covariantly-constant spinors will satisfy the integrability
conditions
\be
(\Gamma_{0 a_i} + J_{a_i b_i}\, \Gamma_{\td 0 b_i})\, \eta=0\,,
\ee
where $J_{a_i b_i}$ are the vielbein components of the K\"ahler form
$J^i$.  It is straightforward to see, generalising the discussion of
section 5.2, that for holomorphic complex 
coordinates on the total space we can use the holomorphic coordinates
of the various factors $M_i$ in the base space ${\cal M}$, together
with the additional complex coordinate $z^0$ given by
\be
z^0 = e^{\rho+\im\, \psi + \sum_i p_i\, K_i}\,,
\ee
where $K_i$ denotes the K\"ahler function on the factor $M_i$, and
$\rho$ is defined by $d\rho = e^{\a_1-2\gamma}\, dr$.  

   Let us present one explicit example of the more general Ricci-flat
K\"ahler solutions:

\bigskip
\noindent\underline{{\bf $\C^2$ bundle over $\CP^1\times \CP^1$}}:
\bigskip

  Consider the case where we take the base space to
be $S^2\times S^2\times S^2$, so $n_1=n_2=n_3=2$.  Then we find
\bea
&&e^{2\a_1} = \ft14 p_1^2\, r^2\,,\qquad e^{2\a_2} =\ft14 p_1\, p_2\,
(r^2 + \ell_2^2)\,,\qquad e^{2\a_3} =
\ft14 p_1\, p_3\, (r^2 + \ell_3^2)\,,\nn\\
&&e^{2\gamma} = \fft{p_1\, k\, r^2\,[\ell_2^2\, \ell_3^2 + \ft23
(\ell_2^2+\ell_3^2)\, r^2 + \ft12 r^4]} {4(r^2+\ell_2^2)\, (r^2+\ell_3^2)}
\,,
\eea
and after making convenient choices $p_i=1$, $\lambda_i=1$ for the
constants, the metric is given by
\be
d\hat s_8^2 =  U^{-1}\,  dr^2 + \ft14 r^2\, U\,
\sigma^2 + \ft14 r^2 \,
d\Omega_1^2 +  \ft14 (r^2+\ell_2^2)\, d\Omega_2^2 + 
 \ft14 (r^2+\ell_3^2)\, d\Omega_3^2 \,,\label{s2s2s2metric}
\ee
where 
\be
U= \fft{3 r^4 + 4(\ell_2^2 + \ell_3^2)\, r^2  + 
              6\ell_2^2\, \ell_3^2}{6(r^2+\ell_2^2)\, (r^2+\ell_3^2)}\,,
\ee
$d\Omega_1^2$, $d\Omega_2^2$ and $d\Omega_3^2$ are metrics on
three unit 2-spheres, and in an obvious notation we have
\be
\sigma = d\psi +  \cos\theta_1\, d\phi_1 +  
  \cos\theta_2\, d\phi_2 +   \cos\theta_3\, d\phi_3 \,,
\ee
where $\psi$ has period $4\pi$.  The metric approaches $\R^4\times
S^2\times S^2$ at $r=0$, with an $S^2\times S^2$ bolt; topologically,
the manifold is a $\C^2$ bundle over $S^2\times S^2$ (or $\CP^1\times
\CP^1$).

\section{More fractional D3-branes and deformed M2-branes}

\subsection{The resolved fractional D3-brane}

\subsubsection{Harmonic 3-form on the $\C^2$ bundle over $\CP^1$}
   
   This is a special case of the construction section 5, in which the
base space is taken to be just $S^2\times S^2$.  It gives a complete
non-compact manifold that provides a ``small resolution'' of the
singular conifold \cite{candel}.  The metric can be written in the
form \cite{zaytse}
\be
ds_6^2 = \fft{r^2+6\ell ^2}{r^2+9 \ell ^2} \, dr^2 + 
\ft19 \Big(\fft{r^2+9\ell^2}{r^2+6 \ell^2}\Big)\,  r^2\, \sigma^2  + 
\ft16 r^2\, d\Omega_2^2 + \ft16 (r^2+ 6 \ell^2)\, d\wtd\Omega_2^2\,,
\label{funsol}
\ee
where
\bea
&&d\Omega_2^2 = d\theta^2 + \sin^2\theta\, d\phi^2 \,,\qquad
d\wtd\Omega_2^2 = d\td\theta^2 + \sin^2\td\theta\, d\td\phi^2 \,,\nn\\
&&\sigma = d\psi +\cos\theta\, d\phi + \cos\td\theta\,
d\td\phi\,,\label{oms}
\eea
and $\ell$ is a constant.  The radial coordinate runs from $r=0$ to
$r=\infty$.  Near $r=0$, the metric smoothly approaches flat $R^4$ times
a 2-sphere of radius $\ell$, while at large $r$ the metric describes the
cone with level surfaces that are the $U(1)$ bundle over $S^2\times
S^2$.  (We are using the notation of \cite{zaytse} here; it
corresponds in our notation to taking $p=q=\lambda=\td\lambda=1$, and
then sending $r\longrightarrow \sqrt{\ft23}\, r$ and
$\ell\longrightarrow 2\ell$.) 

   From (\ref{genkah}) we see that a holomorphic basis of 1-forms is
\be
\e^0 = - e^0 + \im\, e^{\td 0}\,,\qquad \ep^1 = e^1 + \im\, e^2\,,\qquad
\ep^2 = e^3 + \im\, e^4\,,\label{holoviel}
\ee
where
\bea
&&e^0= h\, dr\,,\qquad e^{5}= c\,\sigma\,,\qquad e^1= a\,
d\theta\,,\qquad e^2= a\, \sin\theta\, d\phi\,,\nn\\
&&e^3= b\, d\td\theta\,,\qquad e^4=b\, \sin\td\theta\, d\td\phi\,,
\eea
and $a$, $b$ and $c$ and $h$ are the metric coefficients in
(\ref{funsol}), given by
\be
a^2=\ft16 r^2\,,\qquad b^2=\ft16 (r^2+\ell^2)\,,\qquad
c^2 = \ft19 \Big(\fft{r^2+9\ell^2}{r^2+ 6\ell^2}\Big)\, r^2\,,\qquad
h^2 = \fft{r^2+6\ell^2}{r^2+9\ell^2}\,.
\ee

    It can be useful also to obtain complex coordinates $z^\mu$ compatible
with the complex structure.  Solving the conditions $(\delta_a^b +
\im\, J_a{}^b)\, \del_b\, z^\mu=0$, we are led to the following choice
of complex coordinates,
\be
z_1 = \tan\ft12\theta\, e^{\im\,\phi}\,,\qquad
z_2 = \tan\ft12\td\theta\, e^{\im\,\td\phi}\,,\qquad
z_3 =\sin\theta\, \sin\td\theta\, e^{\rho - \im\,\psi}
\ee
where $\rho$ is related to $r$ by $3 h^2\, dr= r\, d\rho$, \ie $e^\rho
= r^2\, (r^2+9\ell^2)^{1/2}$.  The
complex vielbein basis given in (\ref{holoviel}) then takes the form  
\be
\ep^0 = c\, \Big(\cos\theta\, \fft{dz_1}{z_1} + \cos\td\theta\,
\fft{dz_2}{z_2} - \fft{dz_3}{z_3}\Big)\,,\qquad
\ep^1 = a\, \sin\theta\, \fft{dz_1}{z_1}\,,
\qquad
\ep^2 = b\, \sin\td\theta\, \fft{dz_2}{z_2}\,,
\ee
which shows that it is indeed holomorphic.  Note that the complex
coordinate $z_3$ is related to the $z^0$ coordinate of the general
discussion in section 5.2 by simple coordinate transformations.

   There is a complex self-dual 3-form, satisfying ${*G_\3}=\im\, G_\3$, 
given by 
\be
G_\3 = \fft1{c\, a^2}\, (e^5\wedge e^1\wedge e^2 - \im \, 
              e^0\wedge e^3\wedge e^4) -\fft1{c\, b^2}\, 
   (e^5\wedge e^3\wedge e^4 - \im\, e^0 \wedge e^1\wedge e^2)\,,
\ee
From this, it follows that $G_\3$ is given by
\be
G_\3 = -f_1\, \bar\ep^0\wedge (\ep^1\wedge \bar\ep^1 + \ep^2\wedge \bar\ep^2)
   + f_2\, \ep^0\wedge (\ep^1\wedge \bar\ep^1 - \ep^2\wedge
\bar\ep^2)\,,\label{tz3form}
\ee
where
\be
f_1 \equiv \fft1{4c\, a^2} - \fft1{4c \, b^2} \,,\qquad
f_2\equiv \fft1{4c\, a^2} + \fft1{4c\, b^2} \,.
\ee
Thus we see that $G_\3$ in general has $(2,1)$ and $(1,2)$ pieces.  It
would become pure $(2,1)$ if $f_1$ vanished.  This would happen only
if the scale parameter $\ell$ were set to zero, since then $a$ and $b$
become equal.  In this limit, the metric reverts to the original
unresolved conifold.  The $(1,2)$ piece does, of course, go to zero
faster than the $(2,1)$ piece as $r$ tends to infinity in the resolved
metric.  Thus the harmonic 3-form $G_\3$ becomes ``asymptotically
pure'' at large distances.

   This 3-form was used to construct a fractional D3-brane in
\cite{zaytse}.  Owing to the (marginal) non-normalisability of the
3-form at large distance, it follows that the solution has a
logarithmic correction to the D3-brane metric function $H$ at large
proper distance, as in (\ref{d3asymp}).  The solution also has a
repulson type of singularity owing to the non-normalisability of
$G_\3$ at small distance.  In the next subsection, we shall address
the issue of supersymmetry.

\subsubsection{The issue of supersymmetry in the Pando Zayas-Tseytlin 
D3-brane}

    In the general discussions of supersymmetry for fractional D3-branes
in \cite{ganpol,gub}, it is argued that the deformed solution will only
be supersymmetric if the complex self-dual harmonic 3-form is purely of
type $(2,1)$.  In fact, it was argued in \cite{ganpol,gub} that the
self-duality of the 3-form already implied that it could contain only
$(2,1)$ and $(0,3)$ pieces, and in \cite{gub} it was proved that the
presence of a $(0,3)$ term would imply that there would be no
supersymmetry.  Since we have found that the self-dual harmonic 3-form
in the resolved D3-brane solution of \cite{zaytse} has both $(2,1)$ and
$(1,2)$ pieces, it is appropriate first to discuss why the $(1,2)$ piece
can in fact be present.  After that, we shall discuss its implications
for supersymmetry.

   The general statement about the duality of $(p,q)$-forms in
six-dimensional K\"ahler spaces is as follows.  One must distinguish
between $(2,1)$ or $(1,2)$-forms that are perpendicular to the K\"ahler
form, $G_{abc}\, J^{ab}=0$, and those that are parallel, $G_{abc} =
K_{[a}\, J_{bc]}$.  Denoting these by $(p,q)_\perp$ and
$(p,q)_\parallel$, we then have, in an obvious notation,
\bea
&&{*(2,1)_\perp} =\im\, (2,1)_\perp \,,\qquad {*(2,1)_\parallel} =
-\im\, (2,1)_\parallel\,,\nn\\
&&{*(1,2)_\perp} =-\im\, (1,2)_\perp \,,\qquad {*(1,2)_\parallel} =
\im\, (1,2)_\parallel\,, \\
&&{*(0,3)} = \im\, (0,3)\,,\qquad {*(3,0)} = -\im\, (3,0)\,.\nn
\eea
We can indeed verify by inspection of (\ref{tz3form}) that the first
term is of type $(1,2)_\parallel$, and the second term is of type
$(2,1)_\perp$.  This is therefore compatible with the fact that $G_\3$
is self-dual, ${*G_\3} = \im\, G_\3$.\footnote{Note that there would be
no such harmonic form of type $(1,2)_\parallel$ in a compact Calabi-Yau
3-fold since it would require the existence of a harmonic $(0,1)$-form
$K_{\bar\a}$, which is excluded by the fact that the first cohomology
group $H^1(Z)$ vanishes.  However, in a non-compact manifold, where
furthermore the harmonic forms are not being required to be
$L^2$-normalisable, such arguments break down.}

    Now let us turn to the question of supersymmetry.  It is shown in
\cite{ganpol,gub} that in the Majorana basis of \cite{schwarz}, the
criterion for unbroken supersymmetry for fractional D3-branes is that in
addition to the usual requirements of the standard D3-brane, the
harmonic self-dual 3-form should satisfy
\be
G_{abc}\, \Gamma^{abc}\, \eta=0\,,\qquad 
G_{abc}\, \Gamma^{abc}\, \eta^*=0\,,\label{susycon}
\ee
where $\eta$ is covariantly-constant in the six-dimensional Ricci-flat
K\"ahler metric.   The Majorana basis implies that the ten-dimensional
Dirac matrices $\hat \Gamma_A$ with spatial indices are symmetric and
real, while the Dirac matrix with the timelike index is antisymmetric
and real.  (These are the conventions of \cite{schwarz}, modified to
our notation where the metric signature is mostly positive.)  In
terms of a $4+6$ decomposition, we shall have
\be
\hat \Gamma_\mu = \gamma_\mu\otimes \oneone \,,\qquad \hat \Gamma_m = 
\gamma_5 \otimes \Gamma_m\,,
\ee
where $\gamma_5=\ft{\im}{4!}\, \ep^{\mu\nu\rho\sigma}\,
\gamma_{\mu\nu\rho\sigma}$ is antisymmetric and imaginary, and the Dirac
matrices $\Gamma_m$ in the six-dimensional space are also antisymmetric
and imaginary.  We also have that the chirality operator
$\Gamma_7=\ft{\im}{6!}\, \ep^{a_1\cdots a_6}\, \Gamma_{a_1\cdots a_6}$
is imaginary and antisymmetric, while $\hat \Gamma_{11}$ is symmetric
and real.  Note that because $\Gamma_7$ is imaginary in the Majorana
basis, this means that $\eta^*$ has the opposite chirality to $\eta$.

    We can now see that if the harmonic self-dual 3-form is written as
\be
G_\3 = G_\3^{\sst{\rm Re}} + \im\, G_\3^{\sst{\rm Im}}\,,
\ee
where $ G_\3^{\sst{\rm Re}}$ and $ G_\3^{\sst{\rm Im}}$ are both real,
then the criterion for supersymmetry is equivalent to
\be
G^{\sst{\rm Re}}_{abc}\, \Gamma_{abc}\, \eta = 0\,,\qquad
G^{\sst{\rm Im}}_{abc}\, \Gamma_{abc}\, \eta = 0\,.\label{susycon2}
\ee
Expressing the conditions in this form has the advantage that it is
now independent of the choice of basis for the Dirac matrices.  In
particular, substituting (\ref{tz3form}) into (\ref{susycon2}), and
making use of the conditions $\Gamma_{12}\, \eta= \Gamma_{\td 1 \td
2}\, \eta = -\Gamma_{0\td 0}\, \eta$ satisfied by the
covariantly-constant spinor $\eta$ (see \cite{clpres}), we
arrive at the conclusion that the resolved D3-brane solution of
\cite{zaytse}, using the Ricci-flat metric on the $\C^2$ bundle over
$\CP^1$ is not supersymmetric, since $f_1$ is non-zero.  This is 
consistent with the fact that the $(1,2)_\parallel$
piece in $G_\3$ is non-vanishing.  One can also demonstrate the
breaking of supersymmetry by a direct substitution of $G_\3$ into
(\ref{susycon}) in the Majorana basis.

\subsection{Harmonic 4-form for $\C^2/\Z_2$ and $\C^2$ bundles 
over $\CP^2$, and smooth M2-branes}

   Let us now consider examples where the 8-dimensional Ricci-flat
solution is obtained by taking the level surfaces to be the $U(1)$
bundle over $S^2\times \CP^2$.  First, we shall choose the case where
the bolt at $r=0$ is $\CP^2$, so the metric is given by
(\ref{s2cp2met}); by our general arguments in section 2.1, we can
expect that a harmonic 4-form should exist for this manifold.  Of
course in this case there is still an orbifold-type singularity in the
metric near the origin, as we discussed earlier.  Afterwards, we shall
present a completely regular generalisation of the solution.

  Making a natural ansatz for a self-dual harmonic 4-form that is
invariant under the isometry group, we obtain equations that admit the
simple solution
\be
G_\4 = \fft{\ell^2}{(r^2+\ell^2)^3}\, \Big[ e^{2\beta}\, 
 \hat e^0\wedge \hat e^{\td 0} 
\wedge \wtd J 
-2 e^{2\a}\, \hat e^0\wedge \hat e^{\td 0}\wedge J +
e^{2\a+2\beta}\, J\wedge \wtd J - 
e^{4\beta}\, \wtd J\wedge \wtd J \Big]\,,\label{g4tzgen}
\ee
where $J$ is the K\"ahler form (\ie volume form) on $S^2$, and $\wtd J$
is the K\"ahler form on $\CP^2$.  Note that $\ft12\wtd J\wedge \wtd J$ is
the volume form on $\CP^2$.  We therefore find
\be
G_\4^2 = \fft{288\ell^4 }{(r^2+\ell^2)^6}\,,
\ee
from which it follows that the harmonic 4-form $G_\4$ is $L^2$
normalisable.  

    By making a canonical choice for the vielbeins and K\"ahler 
structures on $S^2$ and $\CP^2$, we may write $J=e^1\wedge e^2$, $\wtd
J=\td e^{\td 1}\wedge \td e^{\td 2} + \td e^{\td 3}\wedge \td e^{\td
4}$.    It then follows from (\ref{genkah}) that a holomorphic
vielbein basis for the 8-dimensional metric is
\be
\ep^0 = \hat e^0 + \im\, \hat e^{\td 0}\,,\quad
\ep^1 = \hat e^1 + \im\, \hat e^2\,,\quad
\ep^2= \hat e^{\td 1} + \im\, \hat e^{\td 2}\,,\quad
\ep^3= \hat e^{\td 3} + \im\, \hat e^{\td 4}\,,
\ee
and the K\"ahler form is given by
\be
\hat J = \ft{\im}{2}\, ( \ep^0\wedge \bar\ep^0 +
\ep^1\wedge \bar\ep^1 + \ep^2\wedge \bar\ep^2 + \ep^3\wedge \bar\ep^3)\,.
\ee
The harmonic 4-form (\ref{g4tzgen}) can then be rewritten as
\bea
G_\4 &=& -\fft{\ell^4}{4(r^2+\ell^2)^3}\, 
\Big[\ep^0\wedge \bar\ep^0\wedge \ep^2\wedge \bar\ep^2 
 + \ep^0\wedge \bar\ep^0\wedge \ep^3\wedge \bar\ep^3 
- 2\ep^0\wedge \bar\ep^0\wedge \ep^1\wedge \bar\ep^1 \nn\\
&&
-\ep^1\wedge \bar\ep^1\wedge \ep^2\wedge \bar\ep^2
+ 2 \ep^2\wedge \bar\ep^2\wedge \ep^3\wedge \bar\ep^3 
+\ep^1\wedge \bar\ep^1\wedge \ep^3\wedge \bar\ep^3\Big]  
\,,\label{s2cp2g4}
\eea
which shows that it is a $(2,2)$-form.  Furthermore, it satisfies
$G_{abcd}\, \hat J_{ab}=0$, and so it is perpendicular to the K\"ahler
form.  In the notation we used earlier, it is therefore a 4-form of type
$(2,2)_\perp$.

  Solving the equation (\ref{heqn}) for the function $H$ in the
deformed M2-brane (\ref{m2sol}), we first find that
\be
 r^3\, (3r^4 + 8 \ell^2\, r^2 + 6 \ell^4)
\, H' = \beta + \fft{3\, m^2\,\ell^4\,  (3r^2 +
\ell^2)}{(r^2 + \ell^2)^3}\,.
\ee
If the constant of integration $\beta$ is chosen to be $\beta=-3\, m^2$,
then the solution for $H$ is non-singular at $r=0$.  Explicitly, we find
\be
H = 1 - \fft{3m^2\, (3r^2 + 2\ell^2)}{2\ell^{4}\, (r^2+
\ell^2)^2} +\fft{27 \sqrt2\, m^2}{4 \ell^{6}}\, 
\arctan\Big[ \fft{\sqrt2\, \ell^2}{3r^2+4\ell^2}\Big]
\,.\label{hsols2cp2}
\ee
This tends to a constant at small $r$, and at large $r$ it has the
asymptotic form
\be
H\sim 1 + \fft{m^2}{6 r^6} -
\fft{m^2\, \ell^2 }{3 r^8} + \fft{19 m^2\, \ell^4 }{90 r^{10}}+
\cdots\,.
\ee

    The asymptotic behaviour is best analysed using the proper distance
$\rho$, defined by $e^{\a-\gamma}\, dr = d\rho$, as the radial
coordinate.  The coordinates $\rho$ and $r$ are related by
\be
r \sim \ft1{\sqrt2}\Big(\rho -\fft{2 \ell^2}{3\rho} + 
\fft{8\ell^6}{45\, \rho^5} + \cdots\Big)\,.
\ee
Thus in terms of $\rho$, the function $H$ behaves as
follows in the asymptotic region:
\be
H \sim 1 + \fft{4 m^2}{3 \rho^6} -
\fft{416 m^2\, \ell^4 }{45\rho^{10}} + \cdots\,.
\ee

    As discussed in section 4.2, the condition for supersymmetry of the
deformed M2-brane is that the harmonic 4-form should satisfy
\be
G_{ABCD}\, \Gamma_{BCD}\, \eta=0\,,\label{g4con}
\ee
where $\eta$ is covariantly constant in the 8-dimensional transverse
metric.  From the integrability conditions (\ref{tzgenspin}) for
$\eta$, and the form of the harmonic 4-form (\ref{g4tzgen}), it is
straightforward to show that (\ref{g4con}) is satisfied, and so this
deformed M2-brane solution is supersymmetric.

    As we mentioned earlier, the maximum periodicity
$(\Delta\psi)_{\rm max}=\pi$ of the $\psi$ coordinate
on the $U(1)$ fibres, implied by the requirement of regularity of the
principal orbits, means that near $r=0$ the $U(1)$ bundle over $S^2$
gives $S^3/\Z_2$ rather than $S^3$, and so there is an orbifold
singularity at $r=0$.  This can be avoided if we replace $\CP^2$ by
$S^2\times S^2$, since then $\psi$ can have period $2\pi$ instead.
Since this is just a special case of more general $\C^2$ bundles over
$S^2\times S^2$ that we discuss in the next section, we shall not
consider this further here.

   It is worth noting, however, that there does exist a generalisation
of the metrics on complex bundles over $S^2\times \CP^2$ that avoids
the orbifold singularity.  The solution is obtained, along with wide
classes of more general related examples, in \cite{d2frac}.   The
investigation of these generalisations was motivated by a construction
of a six-dimensional example in \cite{tz2}.  Here, we shall just quote
the result for the new eight-dimensional metric over $S^2\times
\CP^2$, referring to \cite{d2frac} for explicit details.  It is given
by
\be
d\hat s^2 = e^{-2\gamma}\, dr^2 + e^{2\gamma}\, \sigma^2 + 
2(r+ \ell_1^2)\, d\Sigma_1^2 + 3(r+\ell_2^2)\, d\Sigma_2^2\,,
\ee
where as usual $d\Sigma_m^2$ denotes the Fubini-Study metric on the
unit $\CP^m$.  The function $\gamma$ is given by
\be
e^{2\gamma} = \fft{4\ell_1^2\, r\, (r^2 + 3\ell_2^2\, r + 3 \ell_2^4)
             + r^2\, (3r^2 + 8 \ell_2^2\, r + 6\ell_2^2)}{
      3(r+\ell_1^2)(r+\ell_2^2)^2}\,.
\ee
When $\ell_1=0$ this reduces, after simple coordinate transformations,
to the metric given in (\ref{s2cp2met}).  However, when $\ell_1$ is
non-zero the entire $S^2\times \CP^2$ remains uncollapsed at $r=0$.
The regularity conditions at $r=0$ now imply that $\psi$ should have
period $\pi$, which is the same as the value dictated by regularity
of the principal orbits.

   It is easy to repeat the analysis of the harmonic 4-form for this
metric.  One finds that there is again a normalisable self-dual form, given by
\be
G_\4 = \fft{\ell_2^2}{(r^2+\ell_2^2)^3}\, \Big[ e^{2\beta}\, 
 \hat e^0\wedge \hat e^{\td 0} 
\wedge \wtd J 
-2 e^{2\a}\, \hat e^0\wedge \hat e^{\td 0}\wedge J +
e^{2\a+2\beta}\, J\wedge \wtd J - 
e^{4\beta}\, \wtd J\wedge \wtd J \Big]\,,\label{g4tzgen5}
\ee
where $J$ is the K\"ahler form on $\CP^1$, and $\wtd J$
is the K\"ahler form on $\CP^2$.  It is again of type $(2,2)_\perp$,
and so it will give a supersymmetric deformed M2-brane solution.

   The solution for $H$ is now given by
\bea
H&=& c_0 +\fft{12m^2\, (3r + \ell_1^2 + 
2\ell_2^2)}{\ell_2^2\, (\ell^2_2-4\ell_1^1)(r+\ell_2^2)^2} -\nn\\
&&\fft{108m^2}{\ell_2^2\,(\ell_2^2 - 4\ell_1^2)}
\sum_{i=1}^3 \fft{2\ell_1^2\, \log (r-r_i) + r_i\, 
\log(r-r_i)}{9r_i^2 + 8(\ell_1^2 + 2\ell_2^2)\, r_i + 
6\ell_2^2\, (\ell_2^2 + 2\ell_1^2)}
\eea
where $r_i$ are the three roots of the cubic expression in
$e^{2\gamma}$.   The function $H$ approaches a positive constant
at $r=0$, and at large $r$ it becomes
\bea
H &=& c_0 + \fft{4m^2\, (2\ell_1^2 +\ell_2^)}{3\ell_2^2\, r^3} -
\fft{4m^2\, (2\ell_1^2 + \ell_2^2)(\ell_1^2+ 2\ell_2^2)}{3\ell_2^2\,
r^4} +\nn\\ 
&&\fft{4m^2\, (32\ell_1^6 + 72\ell_1^4\, \ell_2^2 + 120 \ell_1^2\,
\ell_2^4 + 19\ell_2^6)}{45\ell_2^2\, r^5}\,.
\eea
In terms of proper distance $\rho$, defined by $e^{-\gamma}\, dr=d\rho$,
we have
\be
H= c_0 + \fft{256m^2}{3\rho^6} - \fft{106496m^2\,\ell_2^2}{45\rho^{10}}
\,.
\ee
Note that when $\ell_1=1/2 \ell_2$, the solution becomes rather simpler:
\be
H = c_0 + \fft{m^2\, (20r^2 + 55\ell_2^2\, r + 23\ell_2^4)}{10
(r + \ell_2^2)^5}\,.
\ee

\subsection{Harmonic 4-form for $\C^2$ bundle over $\CP^1\times \CP^1$, and
smooth M2-brane}

   We can also construct a harmonic self-dual 4-form for the
8-dimensional metric with the $S^2\times S^2\times S^2$ base space,
which we obtained in (\ref{s2s2s2metric}).  The natural self-dual ansatz is
\bea
G_\4&=& \hat e^0\wedge \hat e^{\td0}\wedge [e^{2\a_1}\, f_1\, \Omega_1 
 + e^{2\a_2}\, f_2\, \Omega_2 + e^{2\a_3}\, f_3\, \Omega_3] \nn\\
&&- e^{2\a_1+2\a_2}\, f_3\, \Omega_1\wedge\Omega_2 
 - e^{2\a_1 + 2\a_3}\, f_2\, \Omega_1\wedge \Omega_3 -
 e^{2\a_2+2\a_3}\, f_1\, \Omega_2\wedge \Omega_3\,.\label{new4ans}
\eea
If we let $x\equiv r^2$, then the equations that follow from $dG_\4=0$
are
\bea
x\, f_1 + (x+\ell_2^2)\, f_2 &=& \Big( x\, (x+\ell_2^2)\,
f_3\Big)'\,,\nn\\
x\, f_1 + (x+\ell_3^2)\, f_3 &=& \Big( x\, (x+\ell_3^2)\,
f_2\Big)'\,,\\
(x+\ell_2^2)\, f_2 + (x+\ell_3^2)\, f_3 &=&
\Big((x+\ell_2^2)(x+\ell_3^2)\, f_1\Big)'\,,\nn
\eea
where a prime means $d/dx$ here.  

    It is straightforward to solve these equations.  By choosing the
integration constants appropriately we can obtain a solution which
gives a self-dual harmonic 4-form that is normalisable, namely
\be
f_1 = \fft{2 \ell_2^2\, \ell_3^2 + (\ell_2^2 + \ell_3^2)\, r^2}{
              (r^2+\ell_2^2)^2\, (r^2+ \ell_3^2)^2}\,,\quad
f_2 = -\fft{\ell_3^2}{(r^2+\ell_2^2)\, (r^2+\ell_3^2)^2} \,,\quad
f_3 = -\fft{\ell_2^2}{(r^2+\ell_3^2)\, (r^2+\ell_2^2)^2} \,.
\label{newform}
\ee
 
     One can see that in the special case where $\ell_3=\ell_2$, the
solution reduces to the one found in (\ref{s2cp2g4}).  This is not
surprising, since then the final $S^2\times S^2$ factors in $S^2\times
S^2\times S^2$ become an Einstein-K\"ahler 4-manifold, and the 
equations arising from solving for the harmonic 4-form reduce to those
that we had to solve previously for the $S^2\times \CP^2$ base space.

     Using the harmonic 4-form given by (\ref{new4ans}) and
(\ref{newform}), we can construct another completely regular deformed 
M2-brane.  It is easily seen from (\ref{new4ans}) that the magnitude of
$G_\4$ will be given by
\be
|G_\4|^2 = 48(f_1^2 + f_2^2 +f_3^2)\,.
\ee
From the expression (\ref{s2s2s2metric}) for the metric, we find that
$\square\,  H = -\ft1{48}\, m^2 \, |G_\4|^2$ becomes
\be
(\sqrt{g}\, U\, H')' = -\ft1{48}\, m^2\, \sqrt{g} \, |G_\4|^2\,,
\ee
where $\sqrt{g} = r^3(r^2+\ell_2^2)(r^2+\ell_3^2)/128$.  The first
integration gives rise to 
\be
H' = \fft{1}{\sqrt{g}\, U}\Big\{\beta + 
\fft{m^2}{256(\ell_2^2-\ell_3^2)}
\Big[\fft{\ell_2^6}{(r^2 + \ell_2^2)^2} -
\fft{\ell_3^6}{(r^2+ \ell_3)^2}\Big]\Big\}\,.
\ee
The singularity at $r=0$ is avoided by choosing
$\beta=-m^2/256$.  Then we find that the function $H$ is given by
\bea
H &=&  c_0 -
\fft{3m^2\, (\ell_2^2 + \ell_3^2 + 3 r^2)}{2(2\ell_3^2-\ell_2^2)
(2\ell_2^2 - \ell_3^2) (r^2 + \ell_2^2)(r^2 + \ell_3^2)} \label{s23h}\\
 &&
+ \fft{27m^2\,\sqrt{2}}{
4 (2\ell_3^2 - \ell_2^2)^{3/2}\, (2\ell_2^2 - \ell_3^2)^{3/2}}\, 
\arctan\Big[\fft{\sqrt{2(2\ell_3^2 - \ell_2^2)(2\ell_2^2-\ell_3^2)}}
{3r^2 + 2(\ell_2^2 + \ell_3^2)}\Big] \,.\nn
\eea
The coordinate $r$ runs from 0 to infinity, and the function $H$ is finite
and positive definite.  For small $r$, $H$ approaches a constant, and
for large $r$, it behaves as
\be
H\sim c_0 + \fft{m^2}{6r^6} - 
\fft{m^2\,(\ell_2^2 + \ell_3^2)}{6r^8} +
\fft{m^2\, (7\ell_2^4+ 5\ell_2^2\, \ell_3^2 + 7 \ell_3^4)}{
90r^{10}} + \cdots\,.
\ee
As usual, it is helpful to express the asymptotic behaviour in terms of
proper distance $\rho$, defined by $dr/\sqrt{U}=d\rho$.  The $r$ and
$\rho$ coordinates are related, at large $r$, by
\be
r\sim \fft{1}{\sqrt2} \Big( \rho -\fft{\ell_2^2 + \ell_3^2}{3\rho} +
\fft{(\ell_2^2 - \ell_3^2)^2}{6\rho^3} + \cdots\Big)\,.
\ee
In terms of $\rho$, $H$ has the following large-distance behaviour:
\be
H \sim c_0 + \fft{4m^2}{3\rho^6} -\fft{32m^2\, (4\ell_2^4 + 5 \ell_2^2\,
\ell_3^2 + 4\ell_3^4)}{45\rho^{10}} + \cdots\,.
\ee
This deformed M2-brane is therefore completely regular, and it 
has a well-defined ADM mass.  It is again supersymmetric.

    Note that this solution for $H$ reduces to the solution (\ref{hsols2cp2})
if the parameters $\ell_2$ and $\ell_3$ are set equal, as would be
expected in the light of our earlier discussion.  It is interesting
also to note that the solution (\ref{s23h}) becomes especially simple
if the parameters satisfy $\ell_2^2 = 2\ell_3^2$ or $\ell_3^2 =
2\ell_2^2$.  Choosing the first of these two equivalent cases, we then
find that the solution can be written as
\be
H= c_0 + \fft{m^2\, (r^2+ 4\ell_3^2)}{6(r^2+\ell_3^2)\, (r^2+2\ell_3^2)^3}
\,.
\ee

\subsection{Deformed M2-brane on the complex line bundle over $\CP^3$}

    At the end of section 5.3 we described the $2(m+1)$-dimensional
Ricci-flat K\"ahler metrics on the complex line bundles over $\CP^m$,
and we obtained an $L^2$-normalisable (anti)-self-dual harmonic
$(m+1)$-form for each case when $m$ is odd.  In particular, we can
take $m=3$, and consider the 8-dimensional complex line bundle over
$\CP^3$.  The metric is given in (\ref{bbmet}), and the harmonic
4-form can be read off from (\ref{bbharm}).  Equation (\ref{heqn}) for
the M2-brane metric function $H$ can be straightforwardly solved in
this case, giving
\be
H = c_0 + \fft{m^2\, r_0^6}{6\, r^6}\,.
\ee
(We have made an appropriate choice for the normalisation of the
harmonic 4-form.)  Since the radial coordinate $r$ runs from
$r_0$ to infinity, it follows that again we have a completely
non-singular deformed M2-brane. 

    In terms of the proper radial distance $\rho$ defined by
$U^{-1/2}\, dr=d\rho$ for this metric, the asymptotic large-distance
behaviour of the function $H$ in the corresponding resolved M2-brane 
is easily seen to be
\be
H = 1+  \fft{m^2\, r_0^6}{\rho^6} - \fft{m^2\, r_0^{14}}{14 \rho^{14}} 
+\cdots\,.
\ee

   It should be noted that this solution is not supersymmetric.  This
can be shown by substituting $G_\4$ directly into the supersymmetry
condition $G_{ABCD}\, \Gamma_{BCD}\, \eta=0$, and making use of the
integrability conditions (\ref{tzgenspin}), which reduce here to just
$(\Gamma_{0a} + J_{ab}\, \Gamma_{\td0 b})\, \eta=0$.  One finds that the
only solution to all these conditions is $\eta=0$.  Alternatively, we
may observe that although the harmonic $(m+1)$-form constructed in
(\ref{bbharm}) is of type $(\ft12(m+1),\ft12(m+1))$ it is not
perpendicular to the K\"ahler form $\hat J = \hat e^0\wedge \hat e^{\td
0} + r^2\, J$, when the odd integer $m$ is greater than 1.  In
particular, the harmonic 4-form in the complex line bundle over $\CP^3$
is of type $(2,2)$ but does not satisfy $G_{ABCD}\, \hat J_{CD}=0$, and,
as shown in \cite{hawtay}, the vanishing of this quantity is another way
of expressing the criterion for supersymmetry.

\subsection{Deformed M2-brane on an 8-manifold of Spin(7) holonomy}

   Recently a resolved M2-brane was
constructed using a Ricci-flat 8-manifold of Spin(7) holonomy
\cite{clpres}.  We shall summarise the key features of that solution
here, in order to allow a comparison with the resolved M2-branes using
Ricci-flat K\"ahler 8-manifolds (which have $SU(4)$ holonomy) that we
have obtained in this paper.  The metric for the Spin(7) manifold,
which is an $\R^4$ bundle over $S^4$, is
given by
\be
ds_8^2 = \Big(1- \fft{a^{10/3}}{r^{10/3}}\Big)^{-1} \, dr^2 + 
         + \ft{9}{100}\, r^2\, \Big(1- \fft{a^{10/3}}{r^{10/3}}\Big)\, 
         (\sigma_i - A^i)^2 + \ft{9}{20} r^2\, d\Omega_4^2\,,
\ee
where $\sigma_i$ are left-invariant 1-forms on $SU(2)$, $d\Omega_4^2$
is the metric on the unit 4-sphere, and $A^i$ is the
$SU(2)$ Yang-Mills instanton on $S^4$ \cite{brysal,gibpagpop}.  The
Yang-Mills field strengths $F^i_{\a\beta}$ satisfy the algebra of the
imaginary unit quaternions, $F^i_{\a\gamma}\, F^j_{\gamma\beta} =
-\delta_{ij}\, \delta_{\a\beta} + \ep_{ijk}\, F^k_{\a\beta}$.   A
normalisable anti-self-dual harmonic 4-form was found in
\cite{clpres}, with orthonormal components given by
\be
G_{0 ijk} = 6 f\, \ep_{ijk}\,,\quad
G_{\a\beta\gamma\delta} =- 6f\,
\ep_{\a\beta\gamma\delta} \,,\quad
G_{ij\a\beta} = f\, \ep_{ijk}\,
F^k_{\a\beta}\,, \quad G_{0i\a\beta} = -f \, 
F^i_{\a\beta}\,,\label{g4spin7}
\ee
where $f=r^{-14/3}$.  

   The deformed M2-brane is given by (\ref{m2sol}), with \cite{clpres}
\bea
H &=& c_0 -\fft{40000 m^2}{729 a^{16/3}\, r^{2/3} } \Big[9 -
\Big(\fft{a}{r}\Big)^{10/3} +
\fft{3\left( 1 -\fft{a^2}{r^2}\right)}{1- \left(\fft{a}{r}\right)^{10/3}}
\Big]\nn\\
&&+ \fft{32000 \sqrt{2\sqrt5}\, m^2}{243 a^6}\, \Big[(\sqrt5-1)\, 
\arctan\left(\fft{\sqrt5+1+ 4\left(\fft{a}{r}\right)^{10/3}}{
\sqrt{2 \sqrt5(\sqrt5-1)}}\right)\nn\\
&&\qquad\qquad\qquad\qquad +(\sqrt5+1)\, 
\arctan\left(\fft{\sqrt5-1+ 4\left(\fft{a}{r}\right)^{10/3}}{
\sqrt{2 \sqrt5(\sqrt5+1)}}\right) \Big]\,.
\eea
At large $r$, $H$ has the asymptotic form
\be
H = c_0 + \fft{2\,10^5\, m^2}{3^7\, r^6} -
\fft{28\,10^4\, a^{4/3}\, m^2}{2673\, r^{22/3}} + \cdots\,.
\ee
In terms of the proper distance $\rho$, the asymptotic behaviour of the
first two terms in $H$ is the same as in the $r$ coordinate.

   The supersymmetry of the solution was not discussed in \cite{clpres},
but has since been demonstrated in \cite{becker}.  Here, we note that
another simple proof of supersymmetry can be given by making use of the
results in \cite{gibpagpop} on the integrability conditions for the
covariantly-constant spinor in the Spin(7) manifold.  These are all
encapsulated in the equations
\be
4 \Gamma_{0i}\, \eta + F^i_{\a\beta}\, \Gamma_{\a\beta}\, \eta=0\,.
\ee
It useful also to note that these imply other equations, including
\be
\Gamma_{0i}\eta=\ft12 \ep_{ijk}\, \Gamma_{jk}\, \eta\,,\quad
F^i_{\a\beta}\, \Gamma_{0i\beta}\, \eta = 3\Gamma_\a\, \eta\,,\quad 
\ep_{ijk}\, \Gamma_{0ijk} = 6\eta\,.
\ee
Using these equations, and the expressions given in (\ref{g4spin7}) for  
the components of the harmonic 4-form, it is now elementary to verify
that $G_{abcd}\, \Gamma_{bcd}\, \eta=0$, and hence that the single
supersymmetry allowed by the Spin(7) holonomy is preserved in the
deformed solution.

\section{Conclusions and comments on dual field theories}

   The purposes of this paper were manifold.  Our first motivation was
purely formal.  We have provided an explicit construction of self-dual
harmonic forms for a class of complete non-compact Ricci-flat K\" ahler
manifolds in $2(n+1)$ real dimensions.  Specifically, we focused on the
Stenzel metrics \cite{sten}. These spaces have $SO(n+2)$ isometry, with
level surfaces corresponding to $SO(n+2)/SO(n)$ coset spaces.  The
degenerate orbit (``bolt'') corresponds to the base space $S^{n+1}\equiv
SO(n+2)/SO(n+1)$. (The $n=1$ case is the Eguchi-Hanson instanton, and
the $n=2$ case was first constructed by Candelas and de la Ossa \cite{candel}
as the deformed conifold.)  For these manifolds we provided an explicit
construction of all the the harmonic, self-dual, middle dimension forms.
Specifically, the solution for the harmonic $(p,q)$-forms in
$p+q=2(n+1)$ dimensions reduces to finding the solution to two coupled
first-order differential equations, which we solved explicitly.

   Interestingly, the $(p,p)$-form (which implies $n$ is odd) is
proportional to $(\cosh \, r)^{-2p}$ and thus turns out to be
$L^2$-normalisable. On the other hand all the other $(p,q)$-forms (for $n$ odd
or even) are not $L^2$ normalisable, with the degree of divergence
increasing with the value $|p-q|$.

   We also gave a construction of another general set of complete
Ricci-flat metrics, whose homogeneous level surfaces are $U(1)$
bundles over a product of $N$ Einstein-K\" ahler base spaces.  The
regularity of the solution implies that one of the base spaces has to
be $\CP^m$ with its Fubini-Study metric, while the other
Einstein-K\"ahler factors are restricted by topological
considerations.  For example, if they are complex projective spaces
$\CP^{m_i}$, then they must satisfy (\ref{gcdcon}) The total space is
topologically a $\C^{m+1}$ bundle over the remaining base-space
factors.  (The 6-dimensional example where there are just two $S^2$
factors appeared in \cite{candel} and was further discussed in
\cite{zaytse}; the metric has level surfaces that are the 5-manifold
known as $T^{1,1}$, which is a $U(1)$ bundle over $S^2\times S^2$.)
We discussed explicit examples, and constructed normalisable harmonic
4-forms for two 8-dimensional cases, where the base spaces are
$S^2\times \CP^2$ and $S^2\times S^2\times S^2$, and harmonic
$(m+1)$-forms for all the cases with $\CP^m$ as base space, for all
odd $m$.

   These formal constructions of self-dual harmonic forms turn out to
have intriguing applications in the study of deformed $p$-brane
configurations whose transverse spaces are non-compact Ricci-flat
manifolds.  In particular, the fractional $D3$-brane found in
\cite{klebstra} provides the non-singular gravity dual of $N=1$ super
Yang-Mills theory in four dimensions.  A generalisation to a number of
deformed $p$-brane configurations with odd or even dimensional
Ricci-flat transverse spaces was recently given in \cite{clpres}.  The
systematic construction of the middle-dimension harmonic forms for the
Stenzel spaces, as well as the generalisations given in Section 5
allowed us to provide another set of regular gravity solutions
corresponding in particular to deformed M2-branes with 8-dimensional
transverse Ricci-flat spaces.  We constructed two examples using
Ricci-flat K\"ahler 8-manifolds, and in each case the deformed 
M2-branes are supported by $(2,2)$-harmonic forms that are normalisable,
and so the M2-branes are regular everywhere. In both cases, as well as
for the case of the M2-brane on the Spin(7) manifold that was
constructed in \cite{clpres}, the solutions are supersymmetric.  This
should be contrasted with the 6-dimensional Ricci-flat K\"ahler metric
on the $\C^2$ bundle over $\CP^1$, which has a harmonic form with both
$(1,2)$ and $(2,1)$ contributions.  Consequently, we show that the
fractional D3-brane using this metric is not supersymmetric.

    The deformed M2-branes that we constructed in this paper, and
the previously-known fractional D3-branes, provide supergravity duals
to field theories with less than maximal supersymmetry.  In fact, the
lower-dimensional conformal symmetry associated with the AdS/CFT
correspondence can be broken by the extra contributions to the
``harmonic'' function $H$ of these resolved branes.  Indeed, in all
the known fractional D3-branes the function $H$ has a universal
asymptotic logarithmic modification, given by (\ref{d3asymp}), owing to
the (marginal) non-normalisability of the complex harmonic self-dual
3-forms in six-dimensions.  This implies that the geometry no longer
has an AdS$_5$ background, and consequently the dual four-dimensional
Yang-Mills field theory has no conformal symmetry.  General
mathematical arguments imply that for any six-dimensional Ricci-flat
K\"ahler metric with an asymptotically conical structure, complex
harmonic 3-forms will necessarily be non-normalisable.

    By contrast, deformed M2-branes have a richer structure, with a
larger range of possibilities for the asymptotic behaviour.   At large
distance the modification to $H$ takes the form
\be
H = c_0 + \fft{Q}{\rho^6} \Big(1 + \fft{c}{\rho^\gamma} + \cdots\Big)\,.
\label{hmod}
\ee
For our Ricci-flat K\"ahler examples constructed in this paper
$\gamma$ takes the values $\ft83$, 4 for supersymmetric M2-branes, and
8 for the non-supersymmetric solution, whilst for the Spin(7) example
in \cite{clpres}, which is supersymmetric, we have $\gamma=\ft43$.
(The constant $c$ is negative in all cases.)  Thus in all these
examples we have $\gamma>0$, implying that the breaking of the
conformal symmetry of the 3-dimensional field theories is much milder.
In fact after dropping the constant 1 in the function $H$, the
solutions are all asymptotically AdS$_4\times M_7$ at large
$r$.\footnote{Similarly, the resolved dyonic string using the
Eguchi-Hanson metric, which was constructed in \cite{clpres}, has
$H\sim c_0 + Q\,\rho^{-2} - c\,\rho^{-6} +\cdots$, in terms of large
proper distance $\rho$.  As a consequence, the solution with $c_0=0$
is also asymptotically AdS$_3$ \cite{clpres}.}

      The resolved M2-brane and dyonic string solutions can reduce on
the compact level surfaces of the transverse spaces to give rise to
domain walls that are asymptotically AdS.  The asymptotically AdS
geometry is supported, from the viewpoint of the dimensionally-reduced
theory, by a non-trivial (and possibly massive) scalar potential that
has a fixed point.  Thus these geometries describe the renormalisation
group flows of the corresponding dual field theories.  However, they are
very different from those associated with continuous distributed brane
configurations \cite{KLT,FGPW,BS,BSI,dist,BBS,clpuni}.  Notably, there
are fewer supersymmetries in our resolved brane solutions than there are
in the distributed brane solutions, which do not break further
supersymmetry.  Furthermore, the solutions we obtained in this paper are
completely free of singularities, while the distributed branes in
general have singularities, including naked ones.  Finally, while the
distributed brane configurations are naturally dual to the Coulomb
branch of the corresponding dual field theory, the resolved M2-branes we
obtained here, which are coincident rather than distributed, are related
to the Higgs branch.

     In \cite{clpres}, a second deformed M2-brane with Spin(7)
holonomy supported by a harmonic 4-form of the opposite duality was
also explicitly constructed.  In this case the 4-form is
non-normalisable at large $r$, and as a consequence, the modification
to the function $H$ in (\ref{hmod}) has a negative value of $\gamma$,
namely $\gamma=-\ft43$.  Thus unlike the deformed M2-branes we discussed
above, this solution will not approach AdS$_4$ spacetime, and the
corresponding three-dimensional field theory dual would have no
conformal symmetry.  An analogous solution with marginally
non-normalisable large-distance behaviour appears to be absent for the
dyonic string with an Eguchi-Hanson transverse space, which is perhaps
consistent with the more central r\^ole of conformal symmetry in two
dimensional field theories.

   In general a deformed $p$-brane solution has a reduced number of
supersymmetries, or none at all.\footnote{In all the examples that we
have studied, turning on the flux from the harmonic form either breaks
all the supersymmetry, or else it preserves all the supersymmetry that
still remains after replacing the flat transverse metric by the more
general complete Ricci-flat metric.}  In order for the solution to be
free of (naked) singularities, the relevant harmonic form has to be
normalisable at small proper distance.  If the harmonic form is also
normalisable at large proper distance, the solution may become
asymptotically AdS in the decoupling limit, describing the
renormalisation group flow of the Higgs branch of the corresponding
less-supersymmetric dual conformal field theory.  If, on the other
hand, the harmonic form is non-normalisable at large distance, then
the correction terms to the function $H$ will break the AdS structure
completely, and the dual field theory will have no conformal symmetry.

    There are clearly open avenues to be investigated along the formal
directions, by constructing harmonic forms not only in the middle
dimension, and for other types of Ricci-flat even-dimensional manifolds,
such as hyper-K\"ahler ones, as well as odd-dimensional ones.  In
particular, the construction of harmonic forms in other than the
middle-dimension may prove to be useful in the study of a larger class
of deformed branes, thus providing gravity dual candidates for a
larger class of models.  Another intriguing question relates to the
many exact Ricci-flat metrics that we obtained in this paper.   It
would be interesting to see how these are related to the general
investigation of the integrability of the Einstein equations for
cohomogeneity one metrics contained in \cite{danwan2}.

\section*{Acknowledgements}

We are grateful to Michael Atiyah and Nigel Hitchin for useful
discussions on $L^2$ harmonic forms, and to Joe Polchinski for
discussions about the resolved D3-brane of \cite{zaytse}.  
We also thank Igor Klebanov for helpful discussions, which, together
with enlightening material in \cite{herkle}, have clarified the 
precise interpretation of fractional and deformed branes.  Accordingly,
we have adjusted the terminology somewhat from an earlier version of
this paper.  M.C. is
grateful to the Rutgers High Energy Theory Group for support and
hospitality.  M.C. is supported in part by DOE grant DE-FG02-95ER40893
and NATO grant 976951; H.L.~is supported in full by DOE grant
DE-FG02-95ER40899; C.N.P.~is supported in part by DOE
DE-FG03-95ER40917.  The work of M.C., G.W.G. and C.N.P. was supported
in part by the programme {\it Supergravity, Superstrings and M-theory}
of the Centre \'Emile Borel of the Institut Henri Poincar\'e, Paris
(UMS 839-CNRS/UPMC).

\end{document}